\documentclass[sigconf]{acmart}
\pdfoutput = 1
\usepackage{fancyhdr}
\usepackage [latin1]{inputenc}
\usepackage[T1]{fontenc}

\usepackage{changes}
\usepackage{ulem}
\usepackage{xcolor}
\usepackage{xspace}
\makeatletter 
\def\bluewave{\bgroup \markoverwith{\lower3.5\p@\hbox{\sixly \textcolor{blue}{\char58}}}\ULon}
\font\sixly=lasy6 
\makeatother 

\usepackage{tabularx} 

\usepackage{graphicx}

\usepackage{booktabs} 

\usepackage{url}

\usepackage{subcaption}
\usepackage{makecell}

\copyrightyear{2018} 
\acmYear{2018} 
\setcopyright{licensedothergov}
\acmConference[IMC '18]{2018 Internet Measurement Conference}{October 31-November 2, 2018}{Boston, MA, USA}
\acmBooktitle{2018 Internet Measurement Conference (IMC '18), October 31-November 2, 2018, Boston, MA, USA}
\acmPrice{15.00}
\acmDOI{10.1145/3278532.3278536}
\acmISBN{978-1-4503-5619-0/18/10}

\newcommand{\etal}{et al.\xspace}

\begin{document}
\title{Multilevel MDA-Lite Paris Traceroute}

\author{Kevin Vermeulen}
\affiliation{%
  \institution{Sorbonne Université}
}
\author{Stephen D. Strowes}
\affiliation{%
  \institution{RIPE NCC}
}
\author{Olivier Fourmaux}
\affiliation{%
  \institution{Sorbonne Université}
}
\author{Timur Friedman}
\affiliation{%
  \institution{Sorbonne Université}
}

\renewcommand{\shortauthors}{K. Vermeulen \etal}

\begin{abstract}
  Since its introduction in 2006-2007, Paris Traceroute and its
  Multipath Detection Algorithm (MDA) have been used to conduct well over a billion 
  IP level multipath route traces from platforms such as M-Lab.
  Unfortunately, the MDA requires a large number of packets in order to trace
  an entire topology of load balanced paths between a source and a
  destination, which makes it undesirable for platforms that otherwise deploy
  Paris Traceroute, such as RIPE Atlas.
  In this paper we present a major update to the Paris Traceroute tool.
  Our contributions are: (1)
  MDA-Lite, an alternative to the MDA that significantly cuts overhead
  while maintaining a low failure probability; (2) Fakeroute, a
  simulator that enables validation of a multipath route tracing
  tool's adherence to its claimed failure probability bounds;
  (3) multilevel multipath route tracing, with, for the first time,
   a Traceroute tool that provides a router-level view of multipath routes; 
   and (4) surveys at both the IP and router levels of
  multipath routing in the Internet, showing, among other things, that
  load balancing topologies have increased in size well beyond what
  has been previously reported as recently as 2016. 
  The data and the software underlying these results are publicly
  available.
\end{abstract}



\maketitle

\thispagestyle{fancy}%

\section{Introduction}\label{intro}

Since its introduction by Van Jacobson in 1988~\cite{jacobson1988traceroute},
Trace\-route has become ubiquitous on both
end-systems and routers for tracing forward paths through the Internet between source and
destination at the IP level. Network operators use it for
troubleshooting; the network measurement community uses it in its
studies; and vast numbers of route traces are executed daily by long
term Internet survey infrastructure such as Ark~\cite{ark},
M-Lab~\cite{mlab, mlab:ccr}, and RIPE Atlas~\cite{RipeAtlas, ripeatlas:ipj}. Two
updates were proposed to Traceroute in 2006-2007 to take into account
the ever-increasing presence of load balancing routers: the Paris
technique~\cite{Augustin:2006:ATA:1177080.1177100,viger2008detection}, 
for tracing a single clean path through load balancers,
and the Multipath Detection Algorithm (MDA)~\cite{augustin2007measuring,veitch:hal-01298261},
for discovering all of the load balanced paths at the IP level
between source and destination. Well over a billion route traces using
the MDA
have been executed by Ark and M-Lab~\cite{mlab:aims}
in the intervening years, and
the Paris technique is used for route tracing on the over 10,000 RIPE
Atlas probes.

A disincentive to deploying the MDA is the network overhead that it requires.
By way of example, suppose
a given hop in a route being traced is evenly load balanced across two interfaces.
If the MDA were to match the overhead of a typical command line Traceroute tool and
send just three probes per hop, the first
probe will find one interface and the subsequent two probes will together have a 25\%
probability of missing the other interface.
In order to bring the probability of failing to discover both interfaces under 1\%, 
a total of eight probes
would need to be sent to that hop.
Even for a single load balanced hop, we must more than double the workload.
To have a high degree of
confidence in full discovery of full load balanced topologies
requires hundreds or even thousands of packets.
Our work is motivated by the aim of minimising this overhead.

This paper makes four contributions that advance the state
of the art for multipath route tracing in the IPv4 Internet. First is
\textit{MDA-Lite} (Sec.~\ref{mda-lite}), a lower overhead alternative
to the MDA that is tailored to the most common load balanced
topologies that we encounter in the Internet. We identify a
characteristic that we call ``diamond uniformity'' that often
holds and that can permit significant probe savings. Second
is \textit{Fakeroute} (Sec.~\ref{fakeroute}), which
validates, to a high degree of confidence, that a software tool's
implementation of its multipath route detection algorithm performs
as intended on a variety of simulated test topologies. Third is
\textit{Multilevel MDA-Lite Paris Traceroute}
(Sec.~\ref{multilevel-route-tracing}), which, for the first time,
integrates router-level view of
multipath routes, into a Traceroute tool. Until now this has only been
done by other tools once route tracing is complete. Fourth, we provide
\textit{new survey results} (Sec.~\ref{surveys}) for multipath routing in the Internet, both at the
IP level, and at the router level. We report load
balancing practices on a scale (up to 96 interfaces at a single hop)
never before described.

Both our code and our survey results are publicly available at
{\small\texttt{https://gitlab.planet-lab.eu/cartography/}}.
\section{MDA-Lite}\label{mda-lite}
The idea behind the MDA-Lite is that we
can take advantage of prior knowledge of what a route trace is likely
to encounter in order to probe more efficiently. 
Experience tells us, and our survey in Sec.~\ref{mda-survey} confirms,
that some multipath route patterns
are frequently encountered in the Internet, whereas others are not.
The MDA-Lite algorithm operates on the assumption
that a topological feature that we call
``uniformity'' will be prevalent and that another feature that we
call ``meshing'' will be uncommon. It includes tests to detect
deviations from these assumptions. 
We detail these two topological features in Sec.~\ref{uniformity-and-meshing}.

\begin{figure*}[t]
\vspace{-2mm}
\begin{center}
\scalebox{1.0}{
	\includegraphics[width=\textwidth]{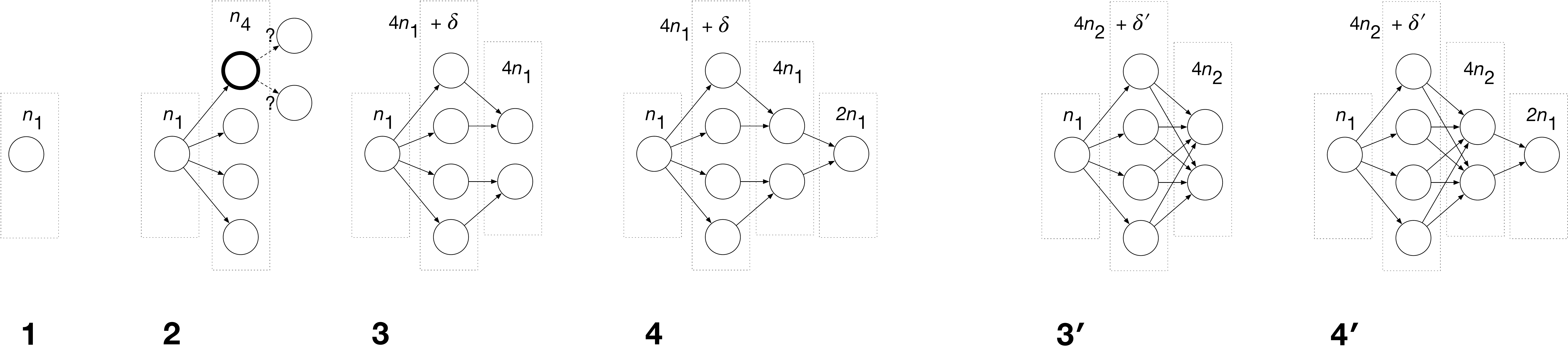}}
	\caption{MDA discovery of an unmeshed and a meshed diamond}
	\label{fig:4-2-lb-step-by-step}
\end{center}
\end{figure*}

\subsection{The MDA and possible probe savings}\label{mda-recap}
This section recalls how the MDA works, stepping us through examples
of the discovery of what are called ``diamonds'', as shown in
Fig.~\ref{fig:4-2-lb-step-by-step}. We see how a feature of the algorithm that we
dub ``node control'' requires large numbers of probes to be
sent. 

The MDA has evolved through 2006 and 2007 poster and workshop
versions~\cite{augustin2006exhaustive, augustin2007multipath} to its
present form in an Infocom 2009 paper~\cite{veitch:hal-01298261}.
This latter publication describes an idealized formal model for
multipath route discovery~\cite[Sec.~II.A]{veitch:hal-01298261}, based
upon a set of assumptions about the Internet, and explains the
adaptations made~\cite[Sec.~III.A]{veitch:hal-01298261}, in crafting
the MDA, to accommodate some divergences assumptions and
reality. These assumptions are: ``(1) No routing changes during the
discovery process. [...]  (2) There is no per-packet load balancing.
(As a result, we can manipulate a probe packet's flow identifier to
cause it to pass through a chosen node.)  (3) Load balancing is
uniform-at-random across successor nodes.  (4) All probes receive a
response.  (5) The effect of sending one probe packet has no bearing
on the result of any subsequent probe. In particular, load balancers
act independently.''

The MDA works on the basis of an \textit{open set} of
vertices~\cite[Sec.~II.A]{veitch:hal-01298261}, each of which has been
discovered but has not yet had its successor vertices identified. A
discovery round consists in choosing a vertex $v$ from the open set
and trying to find all of its successors. Where there is no load
balancing, $v$ has just one successor, but if $v$ is the responding
interface of a load balancing router, there will be two or more
possible successors that can only be identified by stochastic probing.
In the case that concerns us, per-flow load balancing, successors are
found by varying the flow identifier from one probe packet to the
next.  An extension~\cite[Sec.~3.2]{augustin2007measuring}, that we do
not employ here, would allow us to measure per-destination load
balancing, the effects of which are identical to routing insofar as a single
destination is concerned.

The number of probe packets the MDA sends to discover
all successors of a vertex $v$ is governed by a set of
predetermined stopping points, designated $n_k$.  If $k$ successors
to $v$ have been discovered then the MDA keeps sending probes until
either the number of probes equals $n_k$ or an additional successor
has been discovered. In the latter case, the new stopping point
becomes $n_{k+1}$.  Eventually, one of the stopping points will be
reached.
The stopping points are set in such a way as to guarantee that the
probability of failing to discover all of the successors of a given
vertex is bounded. Combined with the assumption of a maximum number
of branching points, this implies a bound on the failure to discover
an entire topology. The MDA takes as a tunable parameter this global
failure probability bound and works backwards to calculate the failure bound
on discovering all the successors to a given vertex, which in turn
determines the values $n_k$.

\textbf{Diamond:} As defined by Augustin et
al.~\cite{Augustin:2011:MMR:2042972.2042989}, a \textit{diamond} is
``a subgraph delimited by a divergence point followed, two or more
hops later, by a convergence point, with the requirement that all
flows from source to destination flow through both
points''. Fig.~\ref{fig:4-2-lb-step-by-step} provides examples of the
MDA successfully discovering the full topologies of two similar
diamonds: each one has a divergence point at hop~1, followed by
four vertices at hop~2, two vertices at hop~3, and a convergence point
at hop~4. Each vertex represents an IP interface, which is to say that
these are IP level graphs, not router-level graphs. The full diamonds
are shown at steps~4 and $4'$ in the figure. We call the one at step~4
an ``unmeshed'' diamond and the one at step~$4'$
``meshed'', the difference relating to the links between hops~2
and 3. Sec.~\ref{uniformity-and-meshing} provides a formal definition of
meshing.
Since discovery is identical for hops~1 and 2, we show the first two steps for the
unmeshed diamond and do not repeat them for the meshed one.  A vertex
at hop~2 of the unmeshed diamond is highlighted and two hypothetical
successors are shown in order to illustrate ``node control'', a
concept described below.

\textbf{Hop 1:} The MDA sends a probe that discovers the single vertex
at hop~1. It continues by sending additional probes to that hop, each
with a different flow ID, until it reaches the stopping point
of $n_1$ probes, at which point it rules out the existence of a second
vertex at that hop. The annotation shows a total of $n_1$ probes
having been sent to hop~1.

\textbf{Hop 2:} The MDA sends a probe that discovers a vertex at
hop~2. As with hop~1, it sends additional probes, each with a different
flow ID, but in this example it discovers a second vertex on or
before having sent $n_1$ probes. Thus the limit becomes $n_2$. Third and fourth
vertices are discovered before $n_2$ and $n_3$, respectively, are met.
When $n_4$ is reached, no fifth vertex has been
found and so the MDA stops scanning this hop.

\textbf{Node control:} When a hop has more than one vertex, the MDA
works on the hypothesis that each of these vertices is a potential
divergence point with successors that are perhaps reachable only via
that vertex. It therefore employs what we dub here \textit{node
  control}, which ensures that each probe packet that goes to the
subsequent hop does so via the chosen vertex.

We have illustrated node control with the highlighted vertex at hop~2,
and the hypothesis that it has two successor vertices at hop~3. The
MDA needs to identify a minimum of $n_1$ flow IDs that bring
probes having a TTL of 2 to the highlighted vertex in order to send
probes to TTL 3 via that vertex. In order to exercise node control for
each of the four vertices at hop~2, a minimum of $4n_1$ probes must be
sent to hop~2. Depending upon the specific stopping point values, it
can be unlikely or even impossible for the $n_4$ probes that had
initially discovered the vertices at hop~2 to have resulted in at
least $n_1$ of them reaching each of the four vertices. To take a
numerical example from Veitch \etal's Table 1~\cite[Sec.~III.B]{veitch:hal-01298261}, 
$n_1 = 9$ and $n_4 = 33$. In this case, it is impossible for the 33 probes that were used
in hop~2 discovery to yield 9 flow IDs for each of hop~2's four
vertices; at least $4 \times 9 = 36$ probes would be required for
that. 36 probes are unlikely to be distributed perfectly evenly,
so some additional probing is necessary.  The annotation at
hop~2 is updated in the illustration for hop~3 to indicate that $4n_1
+ \delta$ probes have been sent to hop~2, where $\delta$ is a
non-negative integer.

The node control problem is an instance of the Multiple Coupon
Collector's problem, which is described by Newman et
al.~\cite{10.2307/2308930} and more recently by Ferrante et
al.~\cite{ferrante2014coupon}.

\textbf{Hop 3:} Having generated the flow IDs necessary for
node control, the MDA now sends probes to hop~3: $n_1$ probes via each
of the four hop~2 vertices. For the unmeshed topology in this example,
only one successor vertex is discovered for each hop~2 vertex.
The annotation shows a total of $4n_1$ probes
having been sent to hop~3.

\textbf{Hop 4:} The MDA also exercises node control at hop~3 in order
to probe hop~4. In this example, since $n_1$ probes have already
reached each hop~3 vertex, no further flow IDs need to be
generated. The annotation shows a total of $2n_1$ probes having been
sent to hop~4, where the diamond's convergence point is discovered.

A total of $11n_1 + \delta$ probes will have been sent overall to
discover this topology. Using the values from Veitch \etal, $99 +
\delta$ probes will have been required by the MDA. The
values from Veitch \etal\ illustrate the cost of node control: $4n_1 = 36$
probes were sent to hop~3, whereas only $n_2 = 17$ probes were strictly
necessary at that hop, and twice as many probes than necessary were sent to hop~4.

\textbf{Hop 2 node control under meshing:} The numbers differ for the
meshed diamond starting at the third hop, which we distinguish in
Fig.~\ref{fig:4-2-lb-step-by-step} with the label $3'$. Each hop~2
vertex has two successors at hop~$3'$, as opposed to just one at
hop~3. Presuming the MDA discovers the second successor in each case,
node control requires additional probes to be sent to hop~2 such that
there are at least $n_2$ flow IDs that reach each vertex at
that hop. The annotation shows a total of $4n_2 + \delta'$ probes
having been sent to hop~2 for the meshed diamond.

\textbf{Hop $\mathbf{3'}$:} As the annotation shows, a total of $4n_2$
probes are sent to hop~$3'$. The meshing results in more probes than
the $4n_1$ probes sent to hop~3 in the unmeshed diamond.

\textbf{Hop $\mathbf{4'}$:} There being only one node at hop~$4'$, the
annotation shows a total of $2n_1$ robes are sent to that hop, just as
for hop~4 in the unmeshed diamond.

A total of $8n_2 + 3n_1 + \delta'$ probes will have been sent overall
to discover the meshed topology. Using the values from Veitch \etal,
$163 + \delta'$ probes will have been required by the MDA. Again, we
see the cost of node control, here accentuated by the multiple
successors to each hop~2 vertex.

\textbf{Per-packet load balancing:} Since per-packet load balancing
was found to be rare in Augustin et al.'s 2011
survey~\cite{Augustin:2011:MMR:2042972.2042989}, we consider that the
assumption (2) of no per-packet load balancing described at the start
of this subsection is a reasonably good one, and we have omitted the
additional packets to check for per-packet load balancing from our
implementation of the MDA, as well as from the MDA-Lite.

\subsection{Uniformity and meshing}\label{uniformity-and-meshing}
As we see in the Fig.~\ref{fig:4-2-lb-step-by-step} examples, the
MDA's use of node control is costly in the number of probes that it
requires. However, node control is only necessary for certain kinds of
diamonds, which we describe here. If diamonds that require node
control are sufficiently rare, an ``MDA-Lite'' could do away with
much of the need for node control. As we shall see, a small degree of
node control is still required in order to determine which sort of
diamond has been encountered. When necessary, the MDA-Lite can
switch over to the MDA with full node control.

We have identified a diamond feature that we call ``uniformity''
that allows full topology discovery without node control. We have also
identified a characteristic of diamonds that we call ``meshing''
that counteracts the potential for probe savings that uniformity
otherwise offers. We define uniformity and meshing here and, as we
show in Sec.~\ref{important-results}, uniform
unmeshed diamonds are indeed very common. Therefore, probe savings can be
realized by using the MDA-Lite.

\textbf{Uniformity:} We define a \textit{uniform hop} as one at which
there is an equal probability for each of its vertices to be reached
by a probe with that hop's TTL and a randomly chosen flow
identifier. For a uniform hop, the failure probability bounds
associated with the MDA's stopping points, the values $n_k$, apply to
discovery of all the vertices at a hop, and node control is not required.
A diamond as a whole is considered a \textit{uniform diamond} if all
of its hops are uniform.

\textbf{Meshing:} As already implied, meshing has to do with the links
between adjacent hops.
Consider hops at TTLs $i$ and $i+1$. We define
these to be \textit{meshed hops} if one of the three following
conditions applies:
\begin{itemize}
\item The hops have identical numbers of vertices and the out-degree of
  at least one of the vertices at hop~$i$ is two or
  more. Equivalently, the in-degree of at least one of the vertices
  at hop~$i+1$ is two or more.
\item Hop~$i$ has fewer vertices than hop~$i+1$ and the in-degree of
  at least one of the vertices at hop~$i+1$ is two or more.
\item Hop~$i$ has more vertices than hop~$i+1$ and the out-degree of
  at least one of the vertices at hop~$i$ is two or more.
\end{itemize}
We define a \textit{meshed diamond} as a diamond with at least one
pair of meshed hops. The right-hand side of
Fig.~\ref{fig:diamond-metrics} illustrates a meshed diamond, in which
hop pairs $(2,3)$ and $(4,5)$ are meshed.

\subsection{The MDA-Lite algorithm}\label{mda-lite-algorithm}

The MDA proceeds vertex by vertex, employing node control to seek the
successors to each vertex individually. The MDA-Lite, however, reserves node
control for particular cases and proceeds hop by hop in the general
case. At each hop it seeks to discover all of the vertices at that
hop, and in doing so discovers some portion of the edges between
that hop and the prior hop. It then seeks out the remaining edges.
It operates on the assumption that the diamonds that it encounters
will be uniform and unmeshed. If this assumption holds, hop-by-hop
probing will maintain the MDA's failure probability bounds. Because
these two topology assumptions might not hold, the MDA-Lite tests for a lack
of uniformity and the presence of meshing using methods that are less
costly than full application of the MDA. When it detects a
diamond that does not adhere to one of the assumptions, it switches to the MDA.
These steps are described below.

\subsubsection{Uniform, unmeshed diamonds}

The MDA-Lite, operating on the assumption that a hop is uniform, sends
probes to that hop without node control. It starts by reusing one flow
identifier from each of the vertices that it has discovered at the
previous hop, continuing with additional previously-used flow
identifiers and then new ones. It applies the MDA's stopping rule to
remain within the MDA's failure probability bounds for vertex
detection.

To take as examples the topologies in
Fig.~\ref{fig:4-2-lb-step-by-step}, the MDA-Lite sends $n_4$ probes to
hop~2, $n_2$ probes to hop~3, and $n_1$ probes to hop~4. Discovery of
all vertices in the diamond therefore requires $n_4 + n_2 + 2n_1$
probes, or 68 probes when applying the values in Veitch \etal's Table~1,
regardless of whether the diamond is unmeshed or meshed. This compares
to the numbers for the MDA that we determined above: $99 + \delta$
probes for the unmeshed diamond and $163 + \delta'$ probes for the
meshed diamond.

Discovering all of the vertices at adjacent hops~$i$ and $i+1$ does
not imply that the MDA-Lite will have discovered all of the
edges. Finishing up the edge discovery is straightforward, though, for
unmeshed hops, in the sense that it is deterministic rather than
stochastic. It consists of tracing backward from each vertex at
hop~$i+1$ that does not yet have an identified predecessor or forward from each vertex at hop~$i$ that does not yet have an identified successor. There are
three cases to consider:
\begin{itemize}
\item Hop~$i+1$ has fewer vertices than
  hop~$i$. For each hop~$i$ vertex that does not yet have an identified successor, the flow
  identifier of a probe that has discovered that vertex is used to
  send a probe to hop~$i+1$. Assuming no meshing, this completes the
  edge discovery.
\item Hop~$i+1$ has more vertices than hop~$i$. For each vertex at hop~$i+1$
  that does not yet have an identified predecessor, the flow
  identifier of a probe that has discovered that vertex is used to
  send a probe to hop~$i$. Assuming no meshing, this completes the
  edge discovery.
\item Hop~$i+1$ has the same number of vertices as hop ~$i$. We apply both of the methods just explained above.
\end{itemize}

Because a diamond could be meshed or non-uniform, the MDA-Lite tries
to detect those cases, as described below.

\subsubsection{Detecting meshing}\label{detecting-meshing}

To detect meshing, stochastic probing is required, and this involves a
limited application of node control. For a pair of hops having two or
more vertices each, the meshing test consists of tracing from the hop
with the greater number of vertices to the one with the lesser number
of vertices, or tracing in either direction if the hops have equal
numbers of vertices. When tracing forwards, meshing is detected if any
predecessor vertex has an out-degree of 2 or more. For backwards
tracing, it is if any successor vertex has an in-degree of 2 or
more. The test requires node control: 
We introduce a parameter, $\phi \geqslant 2$, for the MDA-Lite,
which determines the number of flow
identifiers that have to be generated for each vertex at the hop from which
tracing will begin. Probes with these flow IDs are sent to
the other hop.

The probability of failing to detect meshing depends upon $\phi$. We
calculate this probability as follows. Suppose that the test is
through forward tracing, and let $V$ be the set of two or more
vertices at hop~$i$ and let $\sigma(v)$ designate the set of successor
nodes of a vertex $v \in V$.  When $\phi$ flow IDs are
generated for each vertex $v \in V$ and probes with those flow
identifiers are sent to hop~$i+1$, the probability of failing to
detect meshing is:
\begin{equation}
\prod_{v \in V}{\frac{1}{|\sigma(v)|^{\phi-1}}}
\label{eq:missing-meshing-equation}
\end{equation}
This probability calculation extends with trivial adjustments to the
case of backward tracing.

\begin{figure}[h]
\begin{subfigure}{.25\textwidth}
  \centering
  \includegraphics[width=\linewidth]{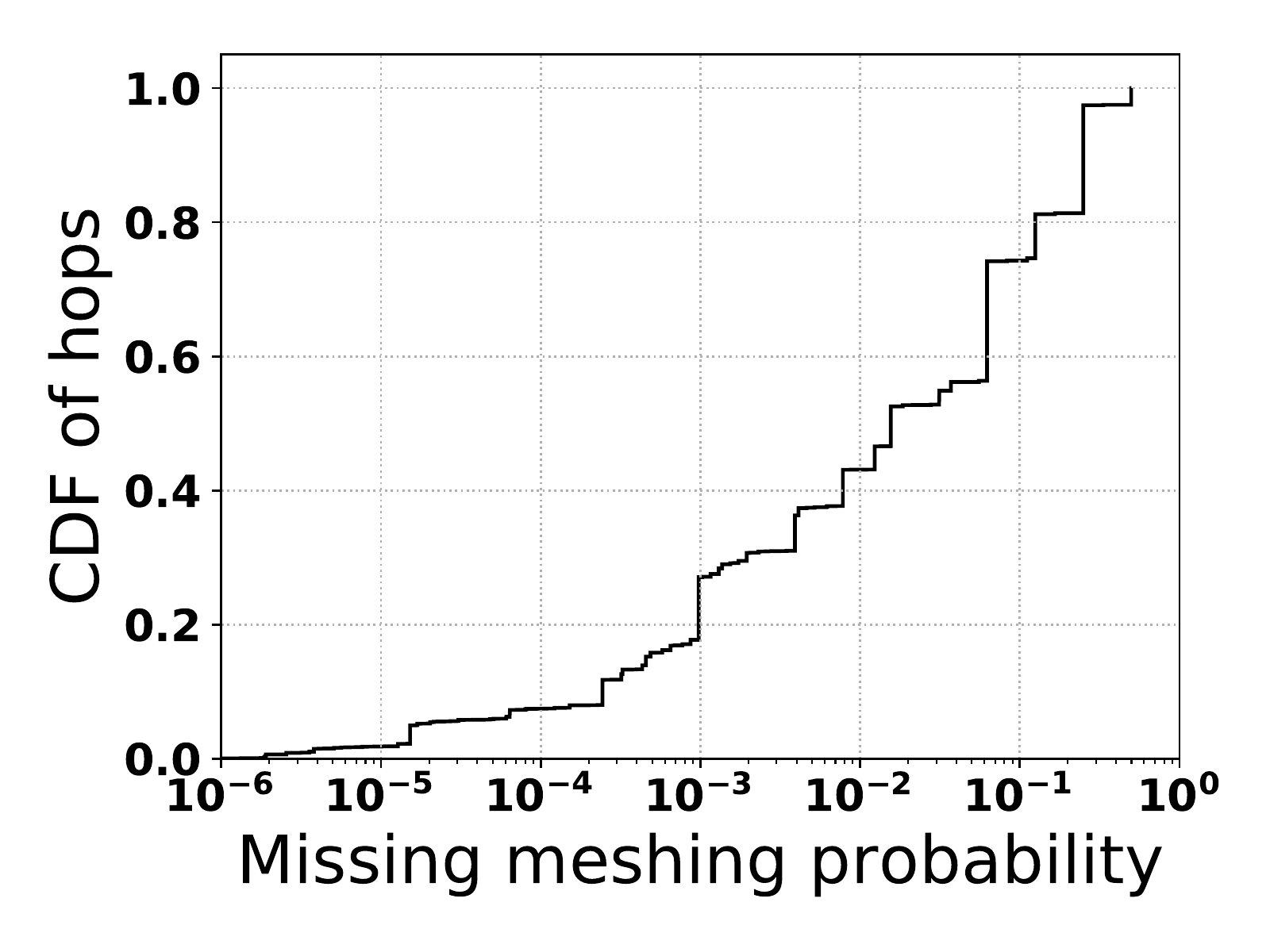}
  \caption{Measured}
\end{subfigure}%
\begin{subfigure}{.25\textwidth}
  \centering
  \includegraphics[width=\linewidth]{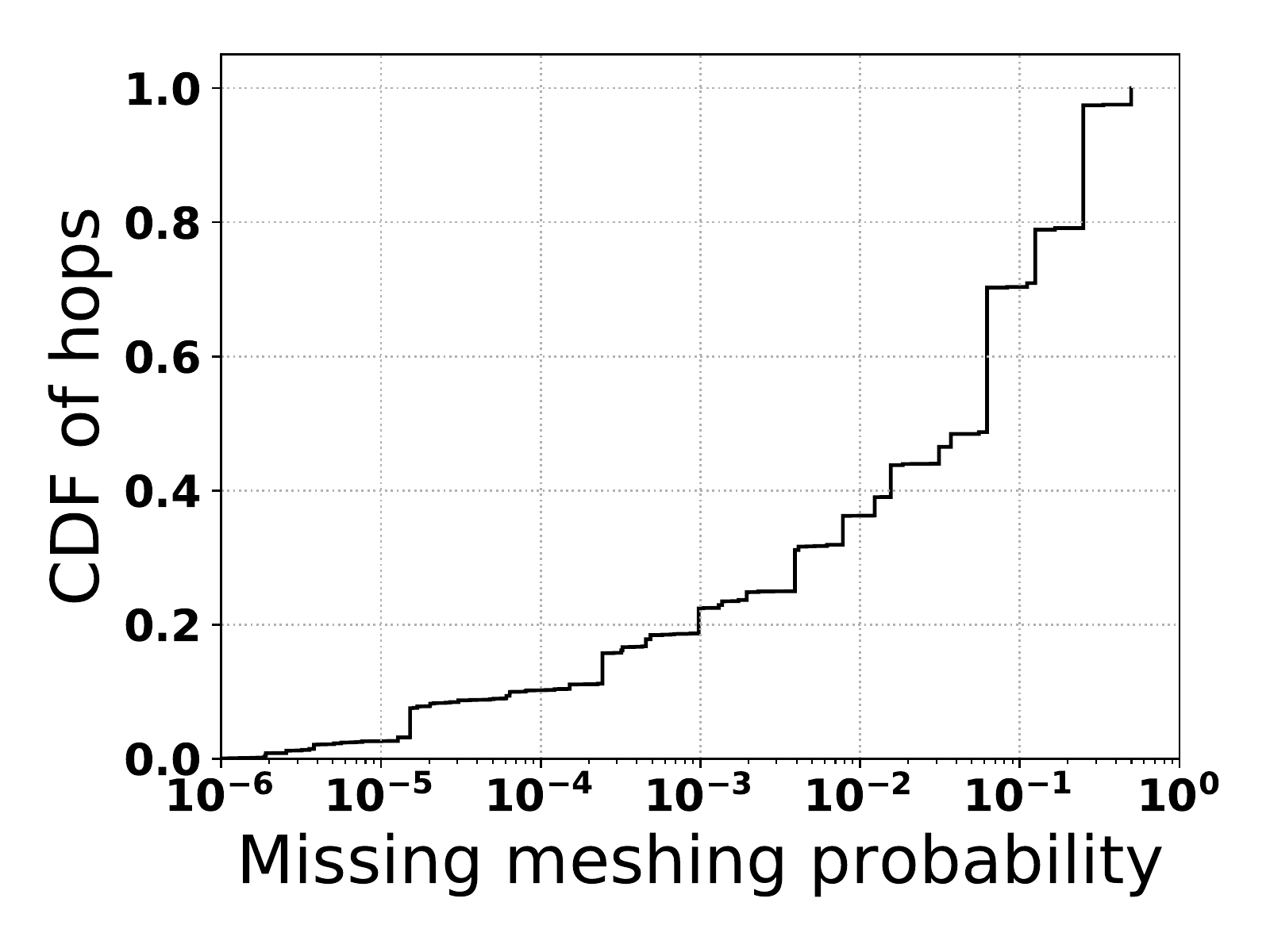}
  \caption{Distinct}
\end{subfigure}
\vspace{-0.5em}
\caption{The probability of failing to detect meshing}
\label{fig:missing-meshing}
\vspace{-1em}
\end{figure}

A minimum value $\phi = 2$ is required in order to detect
meshing. Whether to use a higher value, with a lower failure
probability, is up to the MDA-Lite implementation. We examined how
well this minimum value would work on the meshed diamonds identified by
the MDA in the survey that is described in
Sec.~\ref{mda-survey}. Looking at the topology of each hop pair, we
calculated the probability of the MDA-Lite failing to detect the
meshing. We did this both for \textit{measured diamonds}, which is to say that
each diamond is weighted by the number of times that it is encountered
in the survey, and for \textit{distinct diamonds}, in which we weight each
diamond just once, regardless of how many times it has been seen.
Fig.~\ref{fig:missing-meshing} plots CDFs for the probability of the
MDA-Lite with $\phi = 2$ missing meshing at a hop pair for which the
MDA detected meshing.
We see that, for both measured and distinct diamonds, the probability
of failing to detect meshing is 0.1 or less on 70\% of meshed hop
pairs and 0.25 or less on 95\% of the cases. If we consider this to be too high a probability, $\phi$ is
tunable, and we can set it to 3 or 4.

The overhead generated by the meshing test is lower than the overhead
of the MDA's use of node control. Even with a value of $\phi$ of 3 or
4, this is lower than $n_1 = 9$, the minimum number of flow
identifiers per vertex required by the MDA's use of node control in
Veitch \etal. Furthermore, the MDA-Lite's meshing test is applied only
to a minority of diamonds. As previous surveys have shown, and our
survey confirms, nearly half of all diamonds consist of only a
single multi-vertex hop (48\% for measured and 45\% for distinct diamonds).
The MDA-Lite's meshing test only applies where there are two adjacent
multi-vertex hops, but the MDA applies node control whenever there is
a multi-vertex hop.

\subsubsection{Detecting non-uniformity}

Once edge discovery is complete, and if the MDA has not been engaged
because of meshing, the MDA-Lite tests for non-uniformity. The test is
a purely topological one because the MDA-Lite makes the same
assumption as the MDA about the evenness of load balancing: that each
load balancer dispatches flow IDs in a uniform manner. (Based
upon our experience, this appears to be a realistic assumption, but a
survey on this particular point would be worthwhile.)  What we term
``width asymmetry'' in our survey (see Sec.~\ref{metrics})
is therefore the indicator of non-uniformity.

The MDA-Lite detects width asymmetry as follows. For a pair of
hops~$i$ and $i+1$, if the number of successors is not identical for
every vertex at hop~$i$ or if the number of predecessors is not
identical for every vertex at hop~$i+1$,
the diamond has width asymmetry and is considered to be non-uniform,
and the MDA-Lite switches over to the MDA.

Finding non-uniformity depends upon the topology in question having
been fully revealed. Unlike the MDA, the MDA-Lite does not provide
statistical guarantees on full topology discovery. Rather, the
MDA-Lite assumes that any non-uniformity is likely to be low so
that the full topology will most probably be revealed. 
We empirically justify this assumption based upon our survey results 
in Sec.~\ref{important-results}.

\subsection{MDA-Lite evaluation}\label{mda-lite-evaluation}
\begin{figure}[h]
\begin{subfigure}{.24\textwidth}
  \centering
  \includegraphics[width=\linewidth]{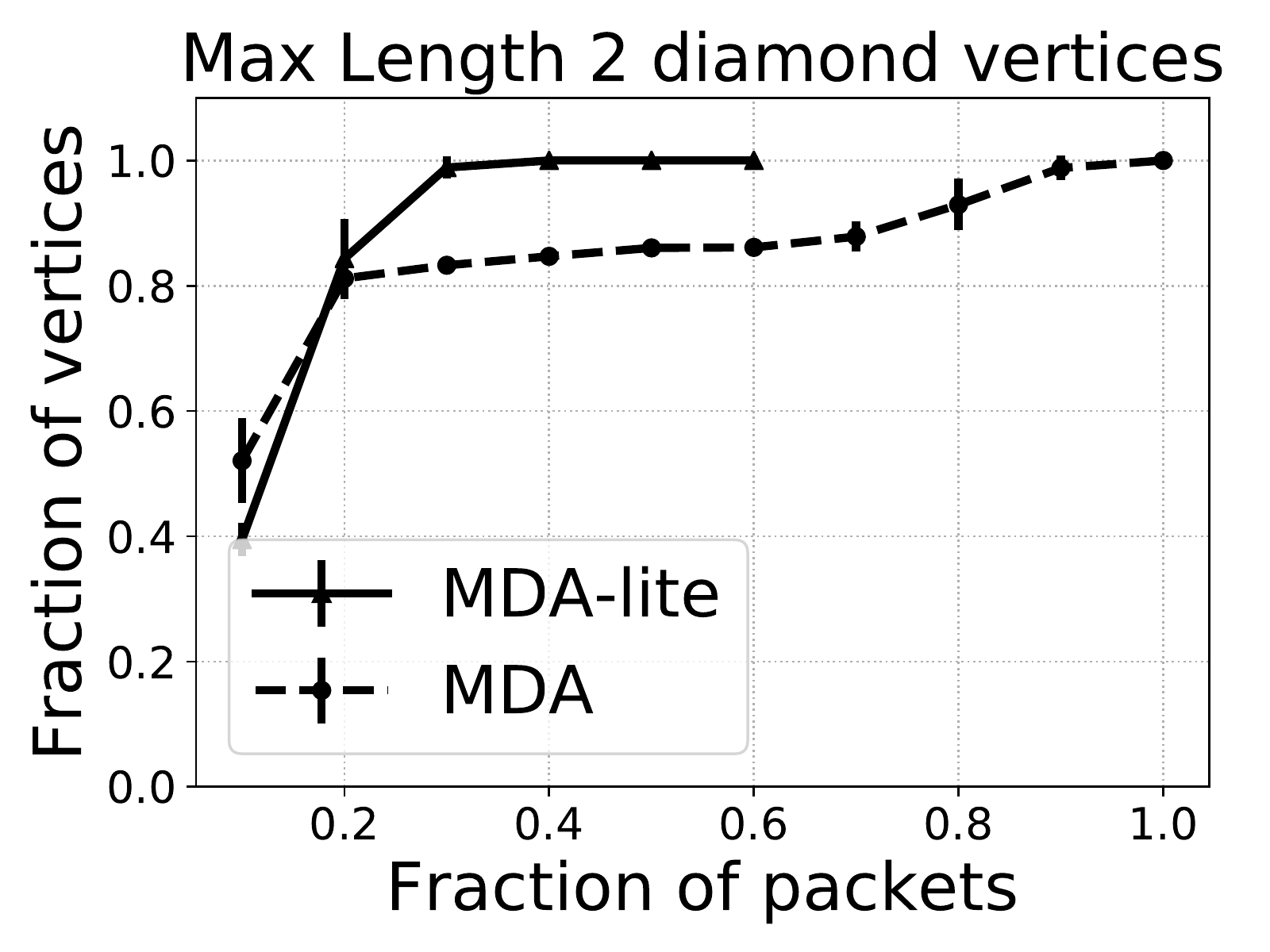}
  \label{fig:sfig1}
\end{subfigure}%
\begin{subfigure}{.24\textwidth}
  \centering
  \includegraphics[width=\linewidth]{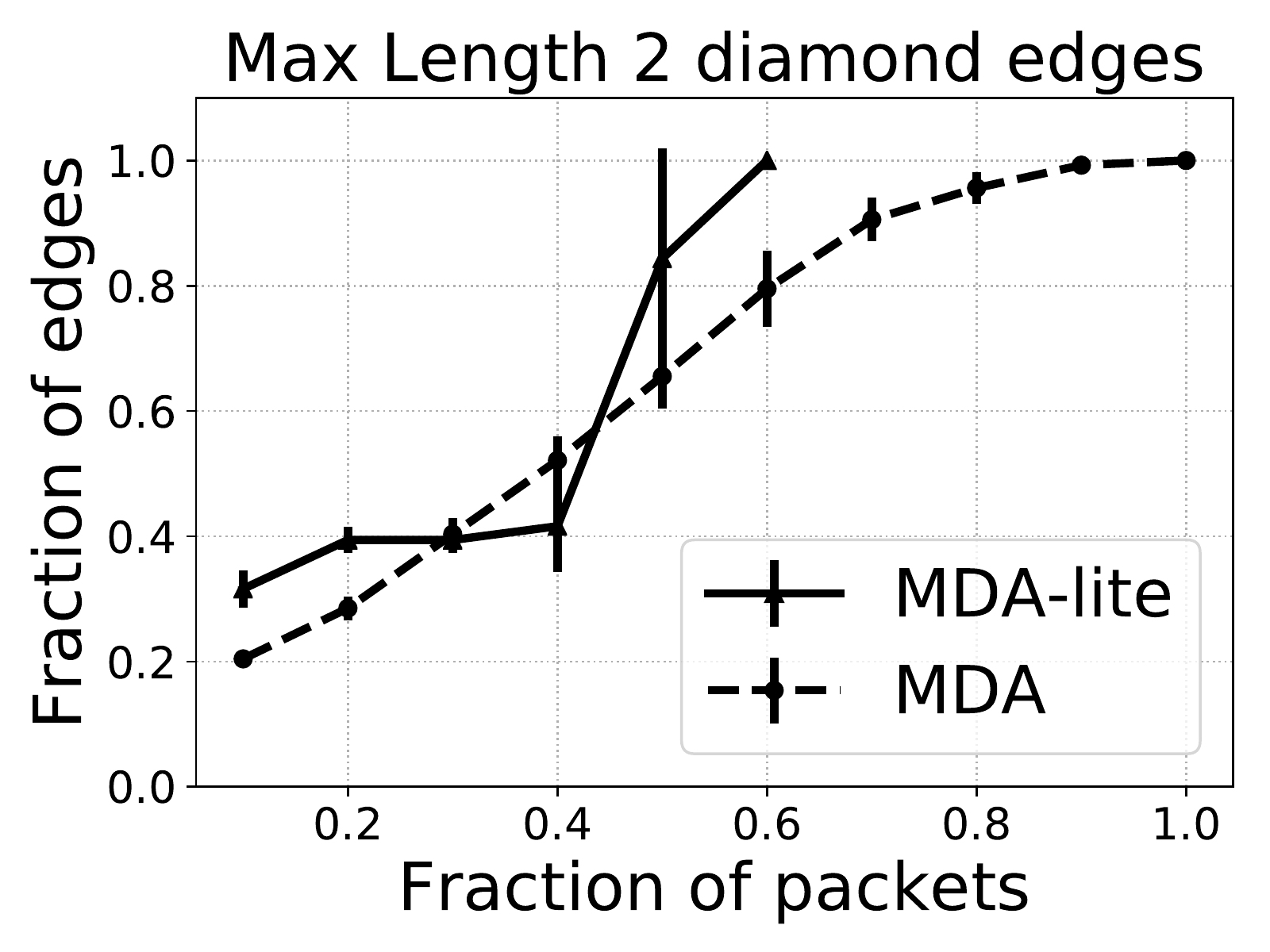}
  \label{fig:sfig2}
\end{subfigure}
\begin{subfigure}{.24\textwidth}
  \centering
  \includegraphics[width=\linewidth]{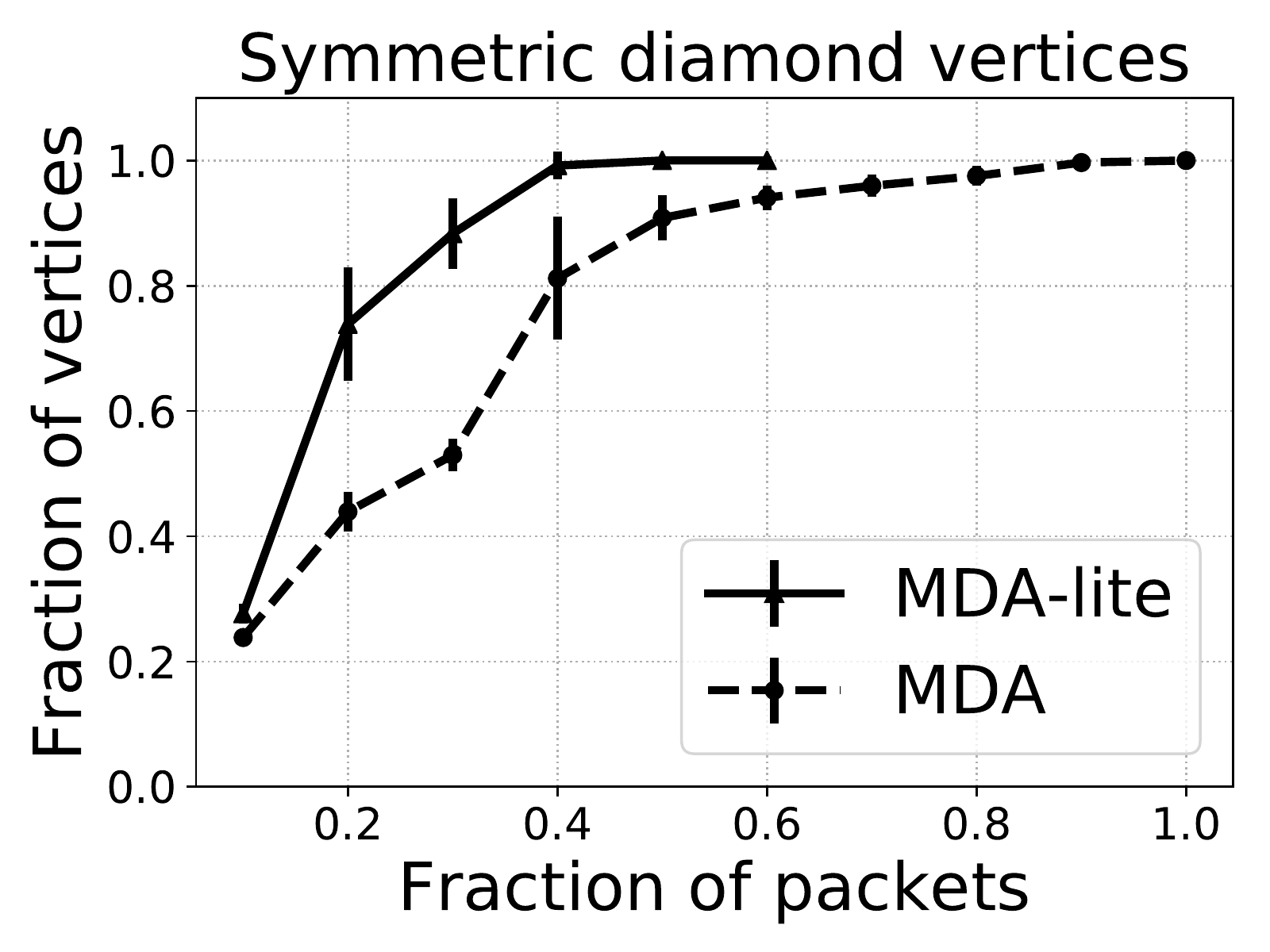}
\end{subfigure}%
\begin{subfigure}{.24\textwidth}
  \centering
  \includegraphics[width=\linewidth]{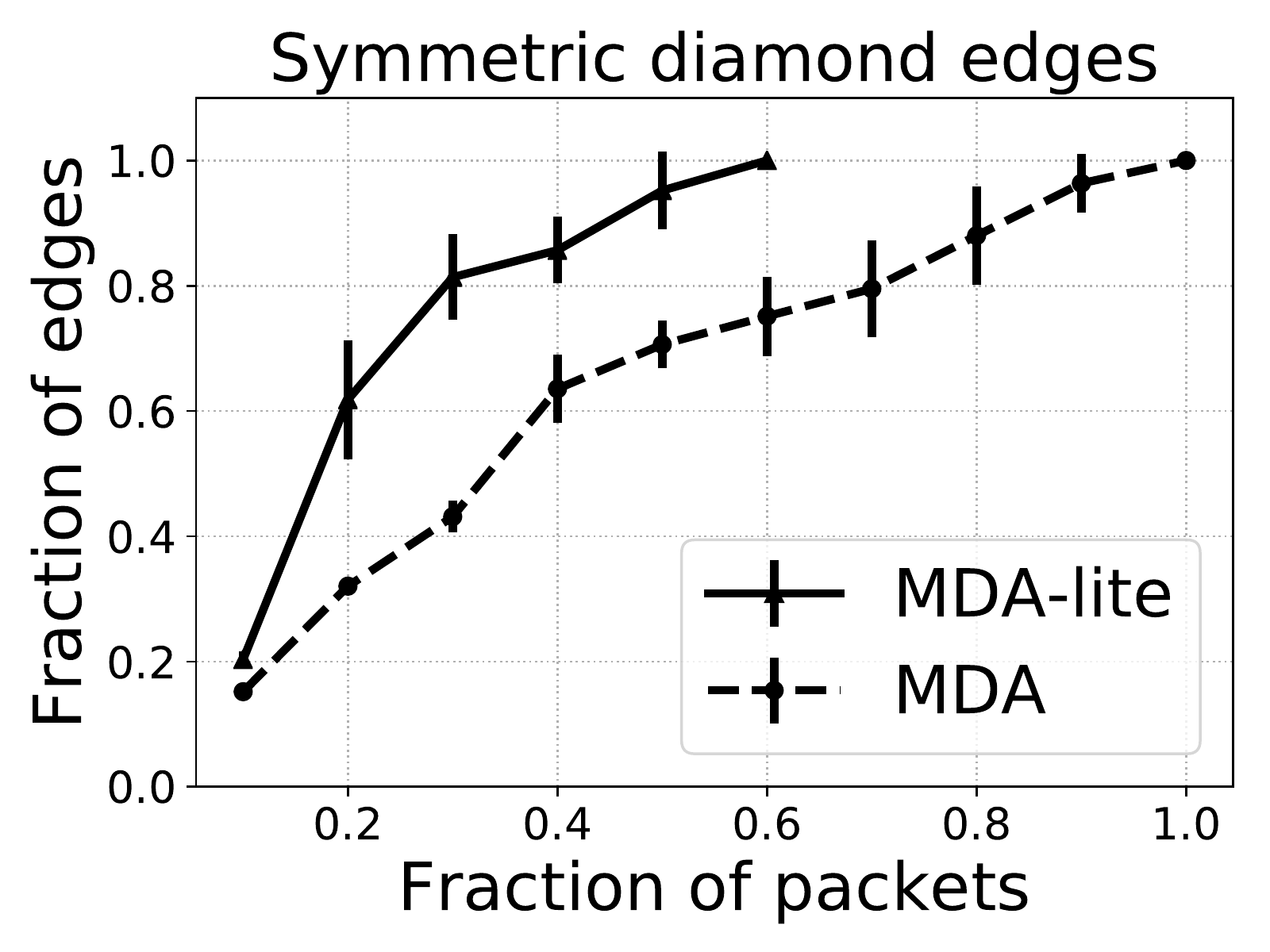}
\end{subfigure}
\begin{subfigure}{.24\textwidth}
  \centering
  \includegraphics[width=\linewidth]{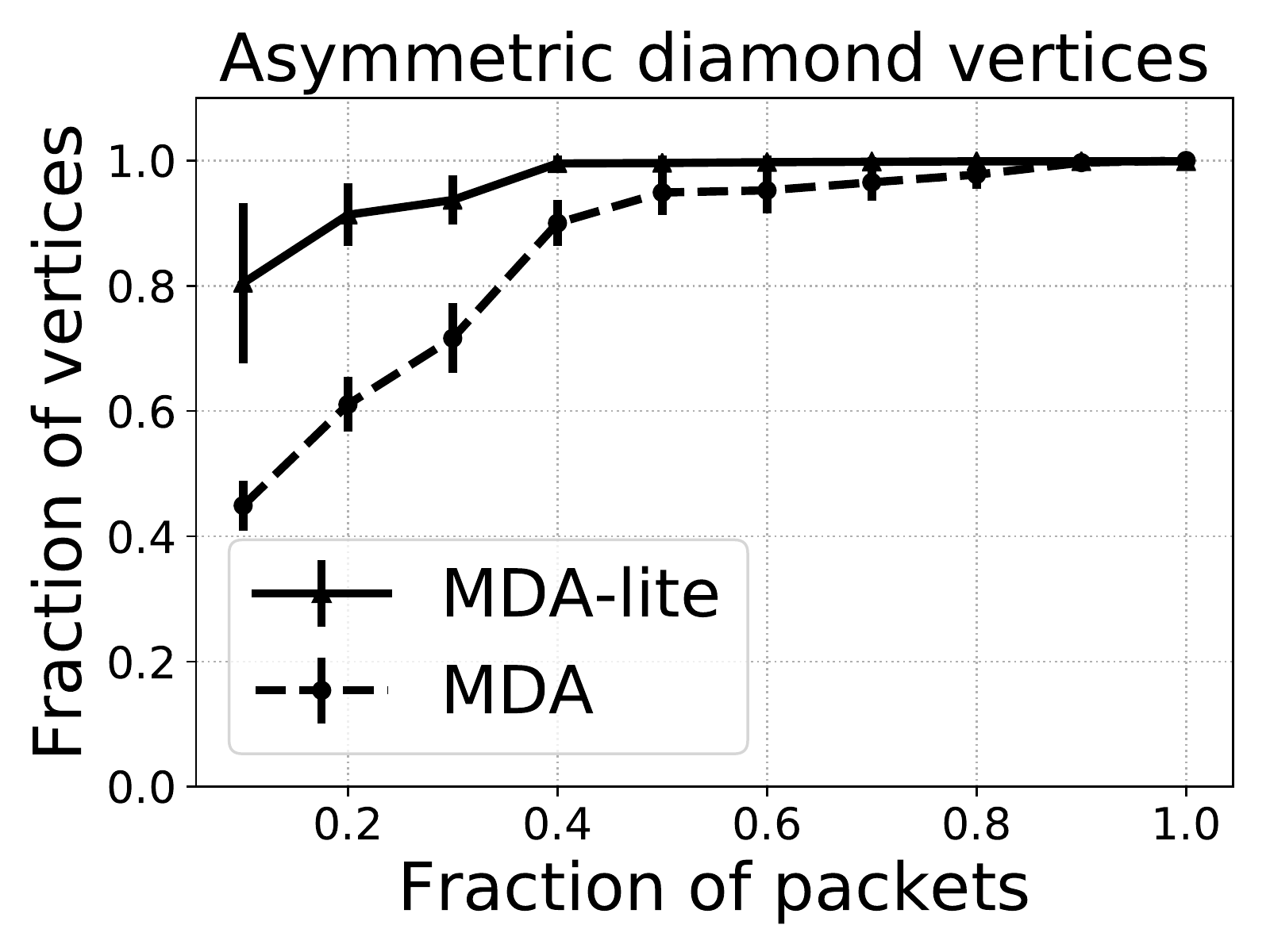}
\end{subfigure}%
\begin{subfigure}{.24\textwidth}
  \centering
  \includegraphics[width=\linewidth]{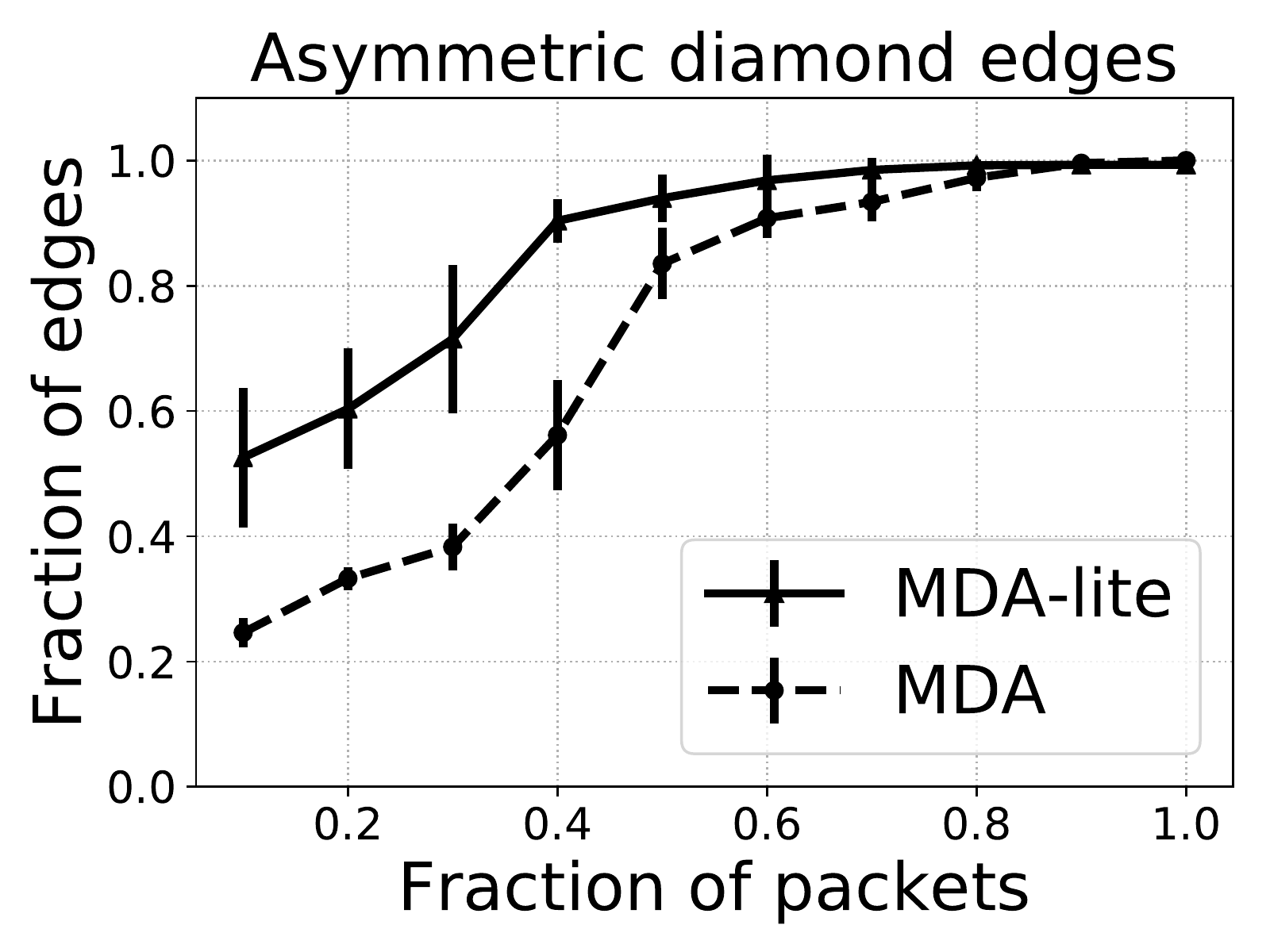}
\end{subfigure}
\begin{subfigure}{.24\textwidth}
  \centering
  \includegraphics[width=\linewidth]{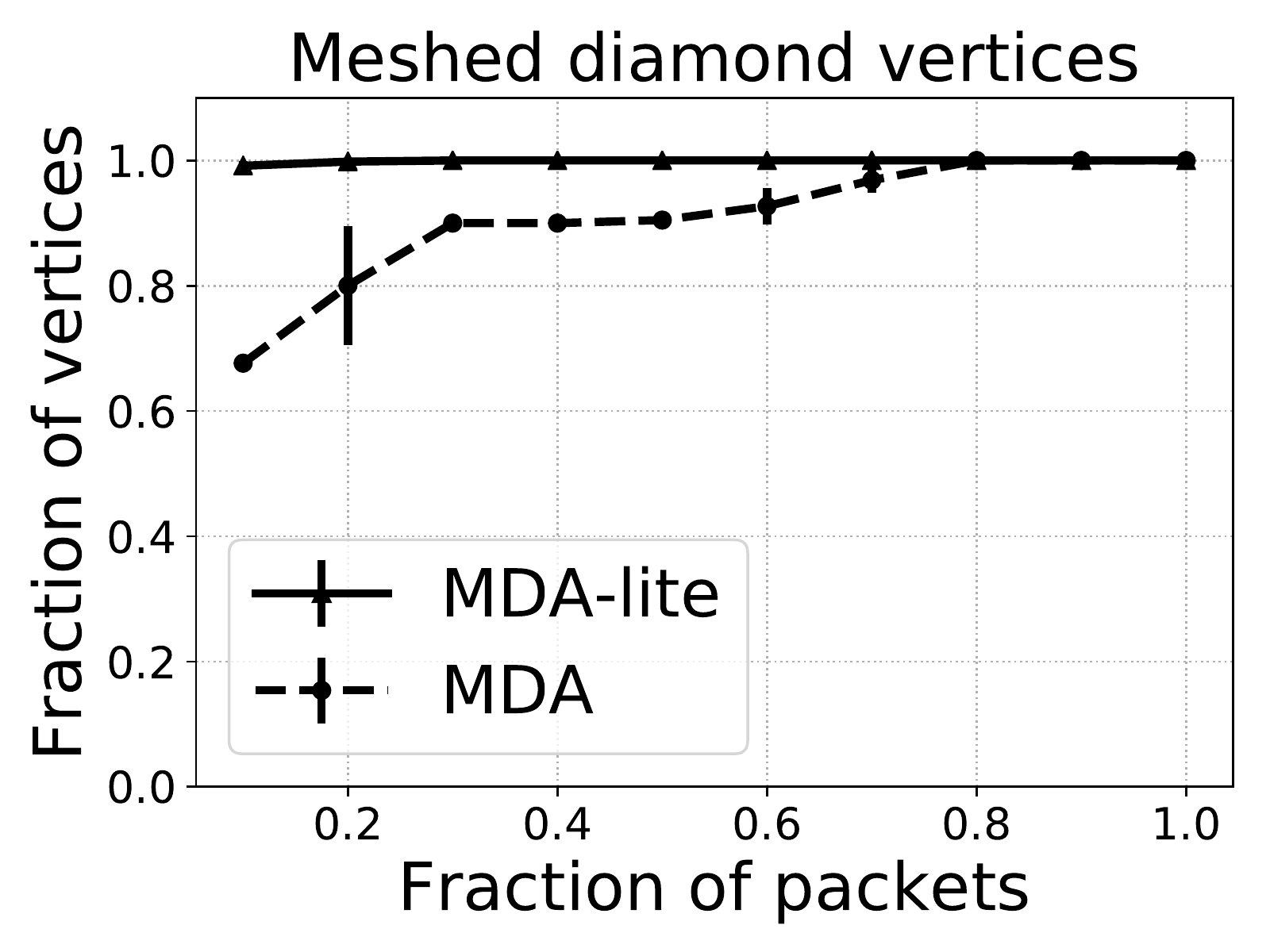}
\end{subfigure}%
\begin{subfigure}{.24\textwidth}
  \centering
  \includegraphics[width=\linewidth]{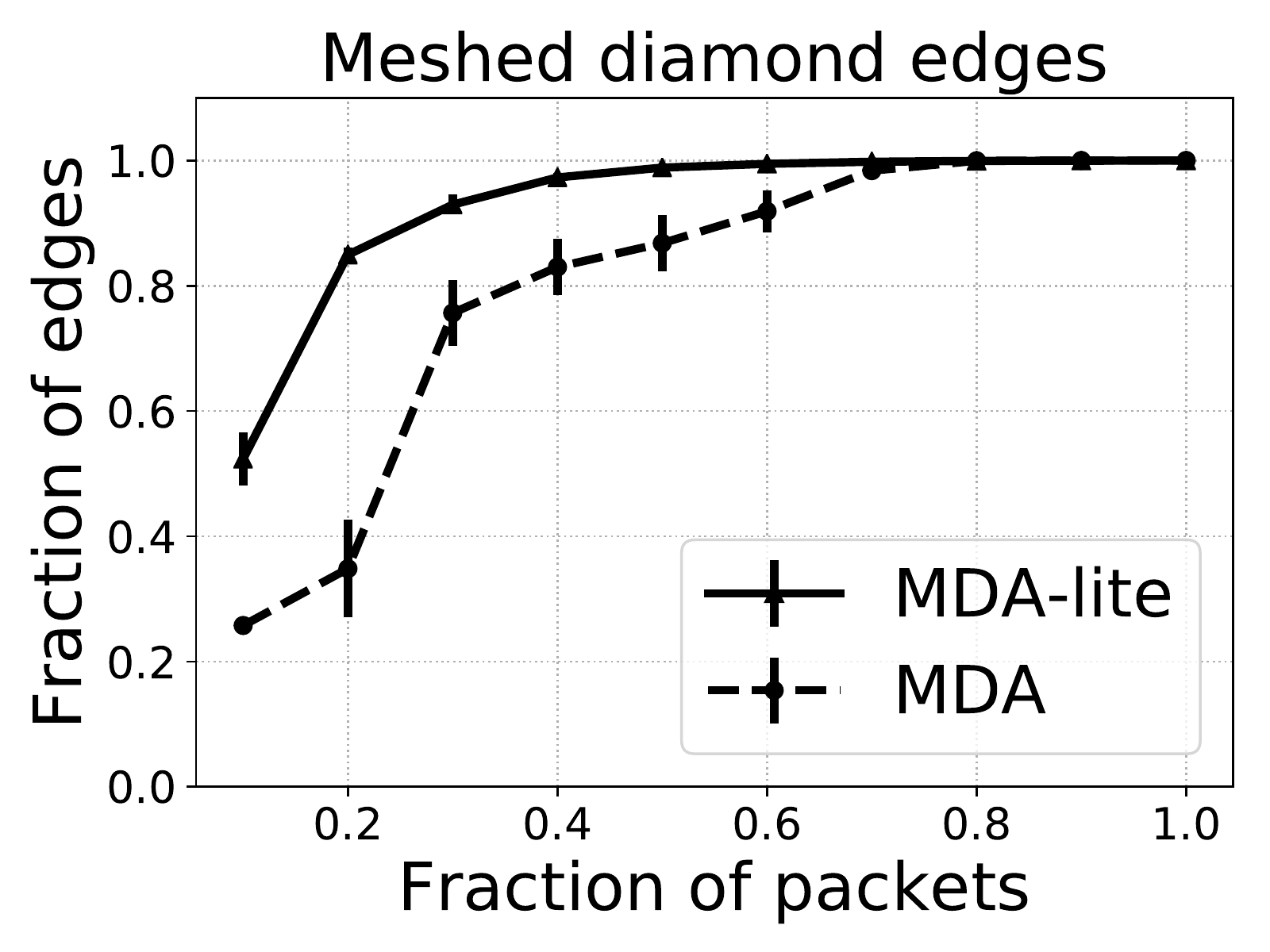}
\end{subfigure}
\caption{MDA-Lite versus MDA simulations}
\label{fig:mda-versus-mda-lite}
\vspace{-2em}
\end{figure}

We have tested the MDA-Lite both through simulations and measurements
on the Internet, finding in both cases that it compares favorably to
the full MDA.
\subsubsection{Evaluation through simulations}\label{mda-lite-simulations}

Simulations allow us to compare the MDA-Lite to the MDA on known
topologies and in an environment free of factors, such as variations
in router load, that are not related to the algorithms. We have chosen
topologies based on both the categories of diamond that are relevant
to the MDA-Lite (uniform or asymmetric, meshed or not, see
Sec.~\ref{mda-lite-algorithm}), and on what we found in our survey
(Sec.~\ref{mda-survey}).
\begin{itemize}
\item The \textit{max length 2 diamond}, found on the trace
  {\small\texttt{pl2.\-prakinf.\-tu-ilmenau.de}} to
  {\small\texttt{83.167.65.184}}, consists of a divergence point, a 28
  vertex hop, and a convergence point. Nearly half of all diamonds in
  the survey are of maximum length 2, though this is a particularly
  wide example. Where the MDA will perform node control on each of the
  28 vertices, the MDA-Lite will avoid doing so. Finding no adjacent
  multi-vertex hops, the MDA-Lite will not apply its meshing test.
\item The \textit{symmetric diamond}, found on the trace
  {\small\texttt{ple1.\-cesnet.cz}} to {\small\texttt{203.195.189.3}},
  has three multi-vertex hops, with 10 being the most vertices at a
  hop. There is no meshing between the hops. On this diamond, the
  MDA-Lite will be obliged to perform a light version of node control
  in order to test for meshing. Finding none, it will not switch over
  to the full MDA.
\item The \textit{asymmetric diamond}, found on the trace
  {\small\texttt{kulcha.\-mimuw.\-edu.\-pl}} to
  {\small\texttt{61.6.250.1}}, has nine multi-vertex hops, with 19
  being the most vertices at a hop. The edges are laid out in such a
  way that at least one of the hops is not uniform, which is to say
  that there is a greater probability of a probe packet with an
  arbitrarily chosen flow identifier reaching some vertices at that hop
  rather than others. It has a ``width asymmetry'' of 17 (this metric
  is defined in Sec.~\ref{metrics}). It is unmeshed. If the MDA-Lite
  discovers the asymmetry, it will be obliged to switch over to the
  full MDA.
\item The \textit{meshed diamond}, found on the trace
  {\small\texttt{ple2.planet\-lab.eu}} to
  {\small\texttt{125.155.82.17}}, has five multi-vertex hops, with 48
  being the most vertices at a hop. It is meshed, and if the MDA-Lite
  discovers the meshing it will be obliged to switch over to the full
  MDA.
\end{itemize}
The simulations ran on Fakeroute, the tool that we describe in
Sec.~\ref{fakeroute}.  Fig.~\ref{fig:mda-versus-mda-lite} shows the
results of 30 runs on each of the four topologies, with vertex
discovery graphs on the left and edge discovery graphs on the
right. Two curves are plotted on each graph: one for the MDA-Lite with
$\phi = 2$ and one for the MDA. The portion of the topologies'
vertices or edges discovered as each algorithm is running is plotted
on the vertical axis. The horizontal axis indicates the number of
probe packets sent, normalized to 1.0 being the number of packets sent
by the MDA in a given run. Since the MDA-Lite sends fewer packets when
confronted with max length 2 and symmetric diamonds, its curves stop
before reaching the right hand side of the graph. Error bars are
given.  We see that the MDA-Lite tends to discover more of these
topologies faster than the MDA, though not always, and that it
discovers the entire topology sooner. In cases where it does not need
to switch over to the full MDA, it also economizes on the number of
probes that it sends, reducing by 40\% the full MDA's overhead on
these examples. For these cases, we see that the MDA-Lite is not
sacrificing the ability to discover the full topology. Because it is
more economical in its use of probes, it discovers more faster.  When
it does not have to switch over to the full MDA, it uses significantly
fewer probes. In the other cases, although it discovers the 
full topology faster than the MDA, the switch to the full MDA means no
economy in its use of probes. 
\begin{figure*}[t]
\vspace{-2mm}
\begin{center}
\scalebox{1}{
  \includegraphics[width=0.33\linewidth]{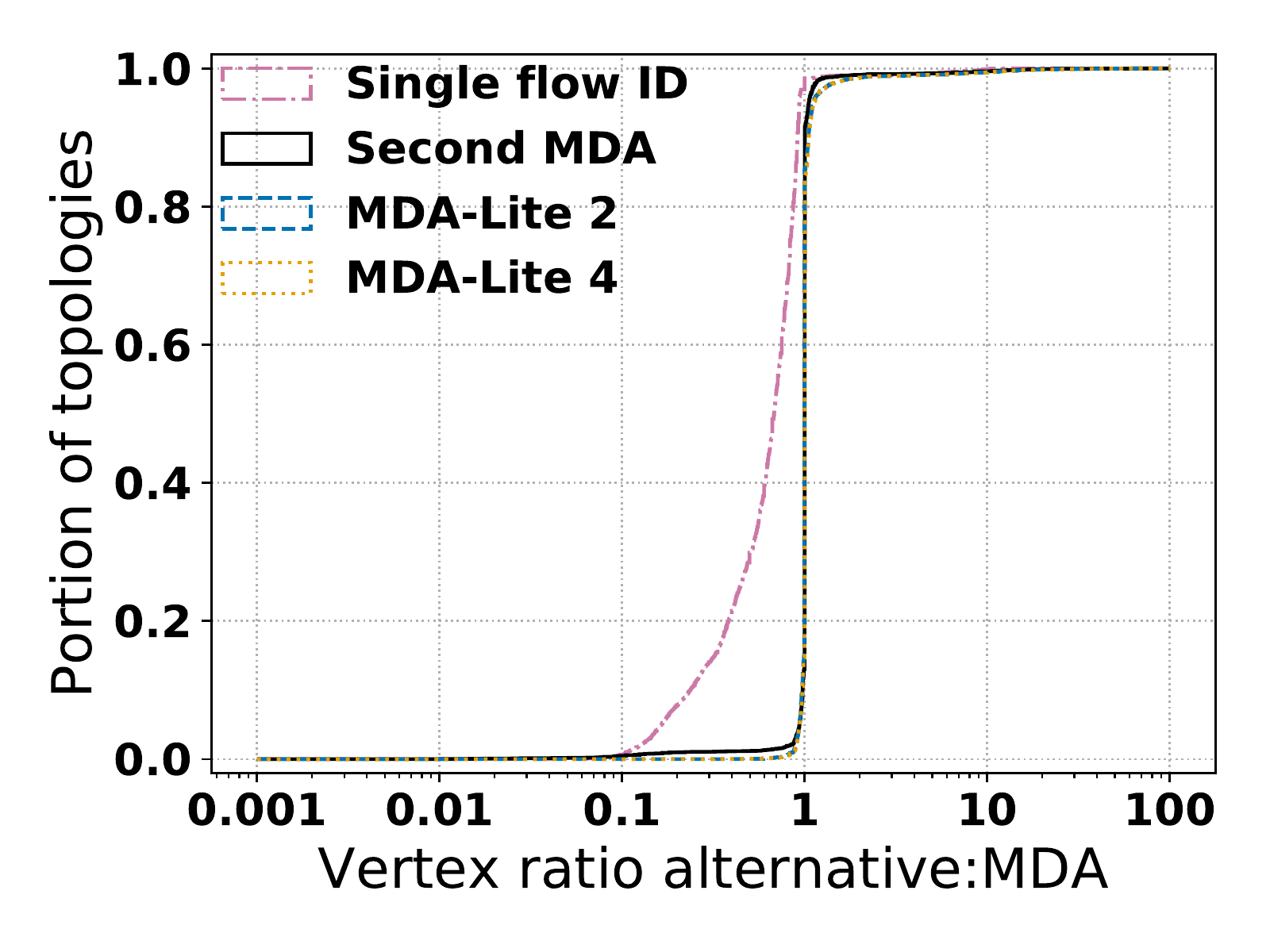}
  \includegraphics[width=0.33\linewidth]{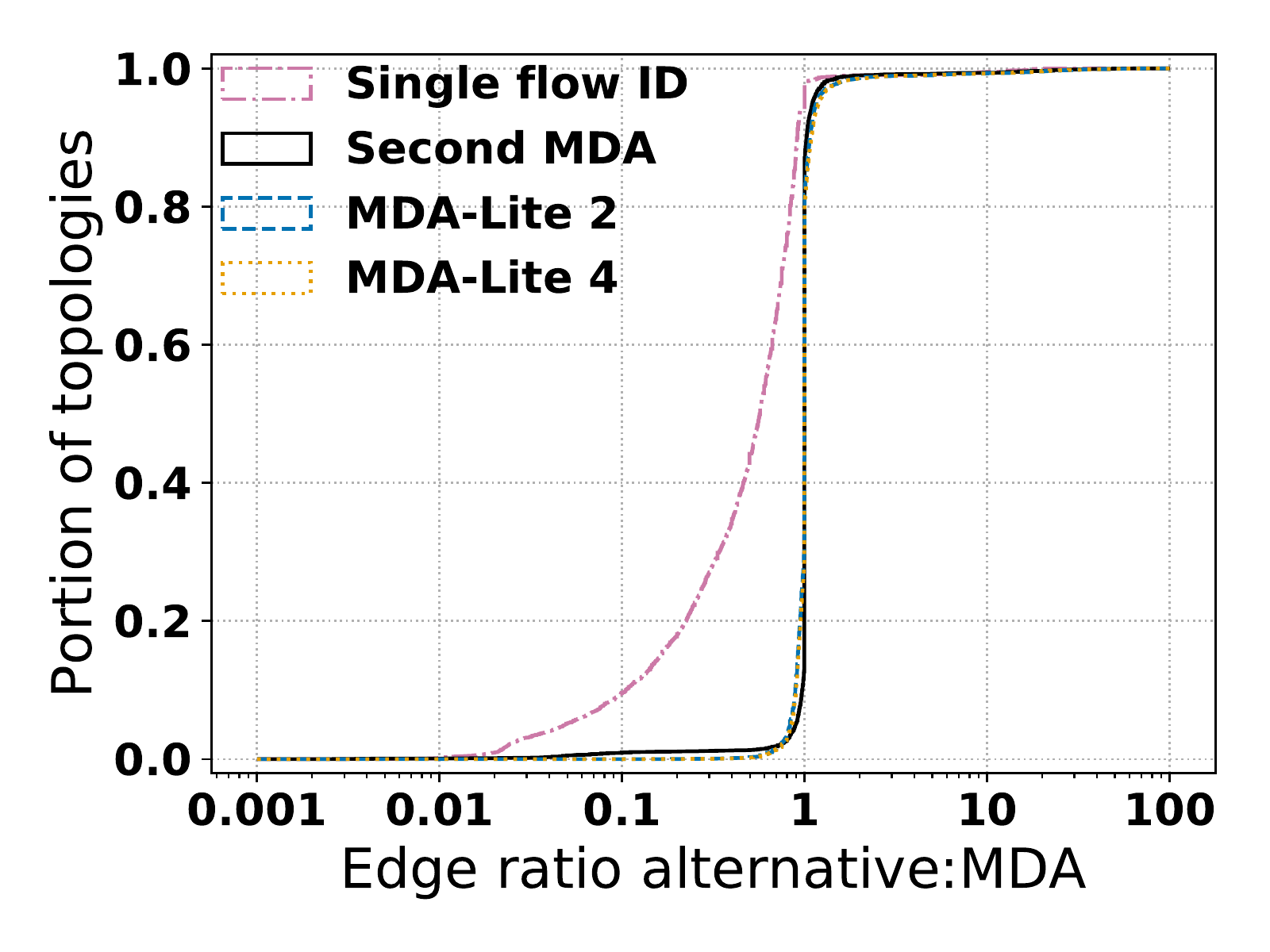}
  \includegraphics[width=0.33\linewidth]{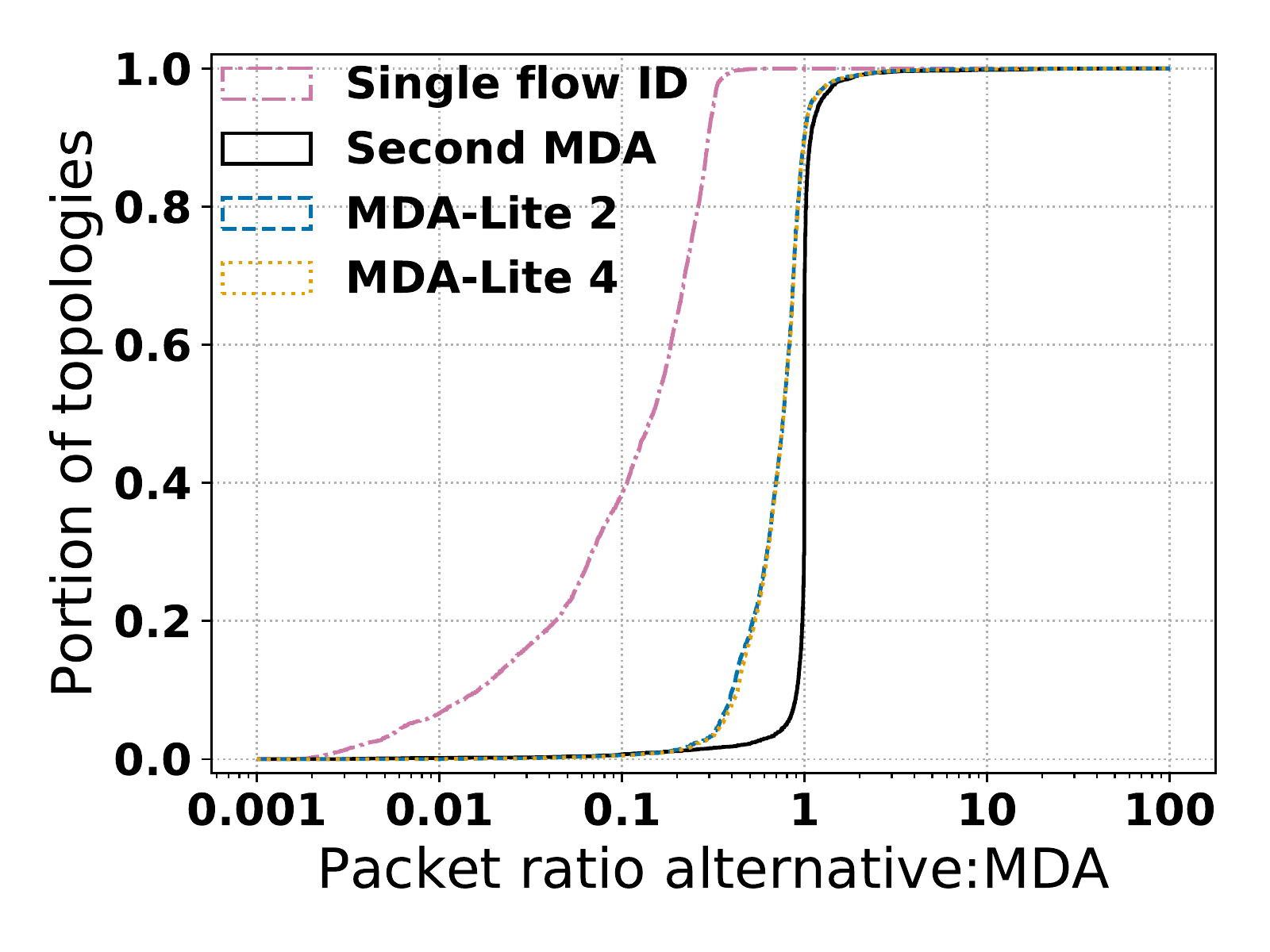}}
\caption{Comparative performance: CDFs over 10,000 measurements in the Internet}
\label{fig:mda-versus-mda-lite-real-world}
\end{center}
\end{figure*}

\subsubsection{Evaluation through measurements}\label{mda-lite-measurements}

We performed our mea\-sure\-ment-based evaluation on a sample of
10,000 source-destination pairs from our survey 
(Sec.~\ref{mda-survey}) for which diamonds had been discovered.
For each of these, we ran five
variants of Paris Traceroute successively: two with the MDA; one with
the MDA-Lite and $\phi=2$; one with the MDA-Lite and $\phi=4$; and one
with just a single flow ID, the way Paris Traceroute is currently
implemented on the RIPE Atlas infrastructure (Sec.~\ref{ripe-mda}).
As a reminder, the parameter $\phi$, defined in
Sec.~\ref{detecting-meshing}, governs how much effort the MDA-Lite
will expend in trying to detect meshing.

For each topology, the first run with the MDA serves as the basis for
comparing the other algorithms. We calculate the ratio of vertices
discovered, edges discovered, and packets sent. The results, plotted
as CDFs, are shown in Fig.~\ref{fig:mda-versus-mda-lite-real-world}.
The horizontal axis plots the ratios in log scale, with $10^0$
indicating that the algorithms performed the same. For the vertex and
edge discovery plots, a value to the left of this value indicates that
the competing algorithm discovered less than the first MDA run, and so
performed worse, and a value to the right indicates that it discovered
more, and so performed better. For the packets plot, at $1$, the
tools sent the same number of packets, whereas a value to the left of
this indicates that the competing algorithm sent fewer packets than
the first MDA run, and so performed better, whereas a value to the
right of this indicates that the competing algorithm sent more packets
than the first MDA run, and so performed worse.

We run the MDA algorithm twice because there are variations from run
to run, both because of changing network conditions and because of the
stochastic nature of MDA and MDA-Lite discovery. The second MDA will
sometimes perform better, sometimes worse than the first, and its
curve, shown as a solid black line in the plots, forms the basis
against which to compare the other algorithms. While the second MDA
performs close to the first, it discovers fewer vertices 12\% of the
time and more vertices 12\% of the time; fewer edges 20\% of the time
and more edges 20\% of the time. We believe that these
differences are largely attributable to the stochastic nature of the
MDA, meaning that either the first or the second run occasionally
terminates its discovery process without having discovered all of the
vertices (and hence edges) that are available to discover. Recall from
Sec.~\ref{mda-lite} that the MDA's failure bound for discovering the
successors to a vertex is set as a function of a global failure bound
for the entire topology and a maximum number of branching points that
the topology might have. This latter parameter is set to 30 by
default, but in complex topologies of the sort that we have
encountered in our survey, there can be far more branching points.

For the comparison between the MDA and the MDA-Lite, we observe that
there is no discernible difference between $\phi=2$ and $\phi=4$ for
the MDA-Lite. Most importantly, the MDA-Lite performs nearly
identically to the second MDA run with respect to the first MDA run:
sometimes better, sometimes worse. Compared to the first MDA run, the
MDA-Lite performed better 14\% of the time and worse
14\% of the time for the vertices; better 20\% of the time
and worse 26\% of the time for the edges. We attribute the larger
number of instances of worse performance to the occasional failure of
MDA-Lite to detect meshing or non-uniformity. The impact of this
greater number on overall performance is negligible, as the ratio
curves are hard to distinguish.

Paris Traceroute with a single flow ID performs notably worse on the
whole than the MDA in both vertex and edge discovery. 
In only 12\% of the cases, we observed at least 90\% of the vertices and
in only 10\% of the cases, we observed at least 90\% of the edges.
We did detect
some outliers where Paris Traceroute with a single flow ID discovers a
greater number of vertices and edges than the MDA. These correspond,
we believe, to cases where the route changed between the runs.

The other aspect of performance that concerns us is the number of
packets that were sent. In 89\% of the comparisons, the MDA-Lite
realized probe savings. We find that we save 40\% of the probes on
30\% of the topologies. The ratio curves for both $\phi=2$ and $\phi=4$ are nearly
identical and they are clearly superior to the curve for the second
MDA run, meaning that when there is a diamond in the topology, the
MDA-Lite will tend to use significantly fewer packets that the MDA.

Paris Traceroute with a single flow ID sends many fewer packets. The
cost of discovering an entire multipath topology via the MDA can be
anywhere from less than 2 times more to 1000 times more than 
the cost of tracing
a single route with a single flow ID.

\begin{table}
\tiny
\begin{center}
\resizebox{\linewidth}{!}{%
\begin{tabular}{|l|r|r|r|}
\cline{2-4}
    \multicolumn{1}{l|}{} & Vertices & Edges & Packets\\
\hline
     MDA 2 & 0.998 & 0.999 & 1.005\\
\hline
	MDA-Lite $\phi=2$ & 1.002 & 1.007 & 0.696\\
\hline
	MDA-Lite $\phi=4$ & 1.004 & 1.005 & 0.711 \\
\hline
	Single flow ID & 0.537 & 0.201 & 0.040\\
\hline
\end{tabular}}
\caption{Comparative performance on aggregated topology:
 ratios with respect to a first MDA round over 10,000 measurements in the Internet
}
\label{table:macroscopic-lite-eval}
\end{center}
\vspace{-3em}
\end{table}

From a macroscopic point of view, Table~\ref{table:macroscopic-lite-eval} 
provides results on the overall topology formed by the aggregation of
the 10,000 measurements of the evaluation dataset. Ratios of topology discovered
and probes sent are computed with respect to the first MDA. 
We see that the topologies discovered by the MDA and the MDA-Lite are very close, 
with a maximum of 0.7\% difference for the edges. We see also that the MDA-Lite
cuts the number of probe packets sent by roughly 30\%.
Paris Traceroute with a single flow ID sends only 4\% of the packets sent by the 
first MDA, but only discovers 53.7\% of the vertices and 20.1\% of the edges. 

\section{Fakeroute}\label{fakeroute}
For any given multipath route between source and destination, one can
calculate the precise probability of the MDA failing to detect the
entire topology. This calculation is a simple application of the MDA's
stopping rule with the chosen stopping points, the values $n_k$
described in Sec.~\ref{mda-recap}, along with the basic assumptions
underlying the MDA, such as that load balancing will be uniform at
random across successor
vertices~\cite[Sec.~II.A]{veitch:hal-01298261}. For a vertex in the
topology that has $K>1$ successors, the first successor will certainly
be found by the first probe packet (among the assumptions is that all
probes receive replies), but there is a probability $1/K^{n_1-1}$ that
a total of $n_1$ probes will fail to discover a second successor,
and the probabilities of failing to discover each of the remaining
successors $k \leqslant K$ are similarly straightforward to
calculate. Veitch et al.\ provide the
details~~\cite[Sec.~II.B]{veitch:hal-01298261}.

In principle, therefore, it should be possible to test that the MDA
has been correctly implemented by a software tool by running it
repeatedly on a suite of benchmark topologies and seeing that the
failure probabilities are as predicted. For scientific purposes, we
would want, if at all possible, to verify the conformance of a tool
before using it, but we have not had that capability until now for
tools that implement the MDA. Our contribution is a network multipath
topology simulator that takes as input a given topology and a number
of values $n_k$ that is at least equal to the highest branching factor
encountered in the topology, that calculates the probability that the
MDA will fail to discover the full topology, and that runs the actual
software tool in question repeatedly on the topology to verify that
the tool does indeed fail at the predicted rate, not more, not less,
providing a confidence interval for this result.

Our Fakeroute is a complete rewrite of the Fakeroute tool that has
been provided as part of the \textit{libparistraceroute}
library~\cite{libparistraceroute}, and which enabled small numbers of
runs of a tool on a simulated topology, for simple debugging purposes,
but that was not designed for large numbers of runs with statistical
validation. The new Fakeroute, written in C++, uses
\textit{libnetfilter-queue}~\cite{libnetfilter-queue} to sniff probe
packets sent by a tool and suck them into the simulated environment
rather than letting them out of the host into Internet. Once a probe
is in, Fakeroute uses \textit{libtins}~\cite{libtins} to read the flow
identifier and TTL from its header fields. These are used to simulate
the probe's passage through the topology, with the pseudo randomness
of load balancing being emulated by the Mersenne
Twister~\cite{mt19937} that comes with the standard C++ library. Using
\textit{libtins}, Fakeroute crafts either an ICMP Time Exceeded or an
ICMP Port Unreachable reply depending on whether the probe is
determined to have reached an intermediate router or the destination,
and sends that back to the tool.  For example, on a topology with the
simplest possible diamond (a divergence point, two nodes, and a
convergence point), we were able to test that the real failure
probability of the topology, which is 0.03125, given the set of $n_k$
values used by the MDA for a failure probability of 0.05, was
respected.  We ran the MDA 1000 times on this topology to obtain a
sample mean rate of failure, and obtained 50 such samples to obtain an
overall mean and a confidence interval. This took 10 minutes on a
contemporary laptop machine, giving a 0.03206 mean of failure, with a
95\% confidence interval of size 0.00156. We were able to run the same test
on much larger topologies as well, as indicated in the previous
section. Fakeroute is available as free open-source software at the
URL mentioned at the end of Sec.~\ref{intro}.

\section{Multilevel Route Tracing}\label{multilevel-route-tracing}
The third principal contribution of this paper, after the MDA-Lite and
Fakeroute of the previous sections, is IPv4 multilevel route tracing,
embodied in a version of Paris Traceroute that we refer to here as
Multilevel MDA-Lite Paris Traceroute (MMLPT).  By ``multilevel'', we
mean that the tool provides router-level information in addition to
the standard interface-level information. Some router-level
information is already commonly provided by standard Traceroute
command line tools, as they perform DNS look-ups on the IP addresses
that they discover, and the name of an interface is often a variant on
the name that has been assigned to the router as a whole. In addition, some of the prior
work~\cite{Vanaubel:2013:NFT:2504730.2504761, barbeiro2016dublin} that
we describe in Sec.~\ref{related-work} can reveal router or middlebox
level information in the context of a Traceroute. Within the network 
measurement community, there are survey workflows, such as the one 
employed by \texttt{bdrmap}~\cite{luckie2016bdrmap}, that perform 
route traces and then alias resolution, and there are survey tools, 
such as \texttt{scamper}~\cite{luckie2010scamper}, that are 
capable of performing both functions independently. To take another 
recent example, Marchetta et al.~\cite{marchetta2016and}, employed a 
specialized tool, Paris Traceroute with the MDA, to conduct multipath 
tracing, and then another specialized tool, \textsc{Midar}, to conduct 
alias resolution on the IP addresses that the first tool reveals. 
But there has not previously been a command-line Traceroute tool, 
in the line of Van Jacobson's Traceroute~\cite{jacobson1988traceroute}, 
Modern Traceroute for Linux~\cite{modern-traceroute-for-linux}, 
and the like, with an option to obtain a router level view of multipath routes.
 With the advent of multipath route tracing ten years ago, it would seem 
 to be a natural next step to incorporate alias resolution directly into Traceroute itself. 
 Such a tool could readily be slotted in to workflows that currently invoke 
 a Traceroute, and it would bring new capabilities to those, such as 
 network operators, who use Traceroute for network troubleshooting purposes.

Alias resolution from a Traceroute perspective, coming as it does from
a single route trace from a single vantage point, will never be as
complete as alias resolution performed from multiple vantage points on
IP addresses gleaned from traces from multiple vantage
points. Nevertheless, we argue, alias resolution integrated into
Traceroute, provides valuable information. When one observes multiple
parallel paths in a route trace, the question immediately arises as to
whether they are independent or not. Between two adjacent hops, one
could be observing links to different interfaces on a single router or
links to separate routers. MMLPT provides the capacity to distinguish
between these cases at the moment of the route trace, without having
to apply an additional tool for post hoc analysis, such Marchetta et al
in~\cite{marchetta2016and}.
Anyone who conducts
route traces outside of the context of a dedicated survey, such as a
network operator performing troubleshooting, can benefit. 

The remainder of this section describes the alias resolution
techniques that MMLPT employs (Sec.~\ref{alias-resolution-impl}) and
shows how we evaluate them (Sec.~\ref{multilevel-evaluation}). Survey
results using the tool are reported in Sec.~\ref{router-survey}.

\subsection{Alias resolution}\label{alias-resolution-impl}

As mentioned in the Related Work section, MMLPT performs alias
resolution using \textsc{Midar}'s Monotonic Bounds Test
(MBT)~\cite{midar} and two techniques described by Vanaubel et al.:
Network Fingerprinting~\cite{Vanaubel:2013:NFT:2504730.2504761}, and
MPLS Labeling~\cite{Vanaubel:2015:MUM:2815675.2815687}.  In its
overall approach, it follows the MBT's set-based schema for alias
identification. An initial set is established of all of the candidate
addresses, and then broken down into smaller and smaller sets as
probing evidence indicates that certain pairs of addresses are not
related. The sets are composed in such a way that each address in a
set has failed alias tests with every address in every other set. At
any point, each set that contains two or more addresses is considered
to consist of the aliases of a common router. Further probing further
refines these sets.

\textsc{Midar} faces a particular challenge in establishing its
initial sets of candidate aliases, as it is designed to seek aliases
from on the order of a million candidate addresses. It breaks this
large number down into manageable sized initial sets by sorting
aliases on the basis of how fast their IP IDs are evolving over
time. 
MMLPT skips this step, as its task is narrower: to seek aliases
among the addresses found in a single multipath route trace. It
assumes that the aliases of a given router are to be found among the
addresses found at a given hop, and so there will be at most on the
order of one hundred candidate aliases. As a result, we only borrow the MBT from
\textsc{Midar}, and not its full complement of probing stages and heuristics.

Evidence that two addresses are not related comes in different forms,
depending upon the test:
\begin{itemize}
\item The MBT looks at sequences of IP IDs from addresses that have
  been probed alternately. A monotonic increase in identifiers, taking
  wraparound into account, is consistent with the addresses being
  aliases, whereas a single out-of-sequence identifier is used to
  place the addresses into separate alias sets.
  We recall that MMLPT has used UDP indirect probing
and that we have used \textsc{Midar} with
UDP, TCP, and ICMP direct probing to collect IP ID time series.
\item Network Fingerprinting looks at the TTLs of reply packets to
  a ping style probe and a Traceroute style probe,
   and infers their likely initial TTLs. Replies to probes of
  different addresses having different initial TTLs 
  are almost
  certainly from different routers, and so the addresses are placed
  into separate alias sets. 
\item MPLS Labeling looks at the MPLS labels that appear in reply
  packets from different addresses. Vanaubel et al.~\cite{Vanaubel:2015:MUM:2815675.2815687}
 have characterized
  the different cases of MPLS tunnels with load balancing and developed 
  methods to infer aliases from MPLS labels.
  To be usable, labels of interfaces in an MPLS tunnel have to be constant 
  over time for each interface. Otherwise, MPLS labels are not helpful to infer
  aliases.
  Then, 
   if, for two interfaces in an MPLS tunnel found at the same hop, 
  their
  labels differ, it is highly likely  that these two
  interfaces belong to two different
  routers. So the addresses are placed into separate alias sets.
  Conversely, if
  the labels are the same for the two interfaces, then it is highly likely that 
  these two interfaces belong to the same router.
\end{itemize}
False positives, in which two addresses that are not aliases remain in
the same set, can arise through their routers having identical fingerprints and
MPLS signatures (when available), alongside a lack of sufficient MBT probing. 
False negatives,
in which two addresses that are in fact aliases get placed in separate
sets can arise when, instead of a single router-wide IP ID counter, a
router employs separate IP ID counters for each flow identifier, and so the
addresses fail the MBT~\cite{chen2005exploiting}. 

Some of the basic data required by these techniques is collected as
part of basic MDA-Lite Paris Traceroute probing: IP IDs that are used
by the MBT; the TTLs of ``indirect probing'' reply packets that are
used by Network Fingerprinting; and the MPLS labels that appear in
reply packets. A light version of the MBT, along with MPLS Labeling,
can therefore be performed ``for free'', based on these data.  The
results are then refined by MMLPT over additional rounds of probing,
with the direct probes required for Network Fingerprinting and
indirect probes to solicit more and longer sequences of IP IDs for the
MBT. The signature-based methods are applied just once, whereas
successive rounds of the MBT refine the results.  After 10 rounds, 
MMLPT declares sets that
remain as aliases.

Our tool, like \textsc{Midar}, produces three possible outcomes for a
pair of IP addresses. Either it accepts that they are aliases of the
same router, or they are rejected as being aliases of the same router,
or it is not possible for the tool to determine one way or the
other. Failure to determine is not an unusual case, as there are
addresses from which responses to probes do not have monotonically
increasing IP ID values. Such an address might, for instance,
systematically respond with the same value in response to every
probe. Or it might not provide a sufficient number of responses from
which to construct a time series.

\subsection{Evaluation}\label{multilevel-evaluation}

\begin{figure}[t]
  \begin{subfigure}{.5\textwidth}
    \centering
    \scalebox{0.7}{
      \includegraphics[width=\linewidth]{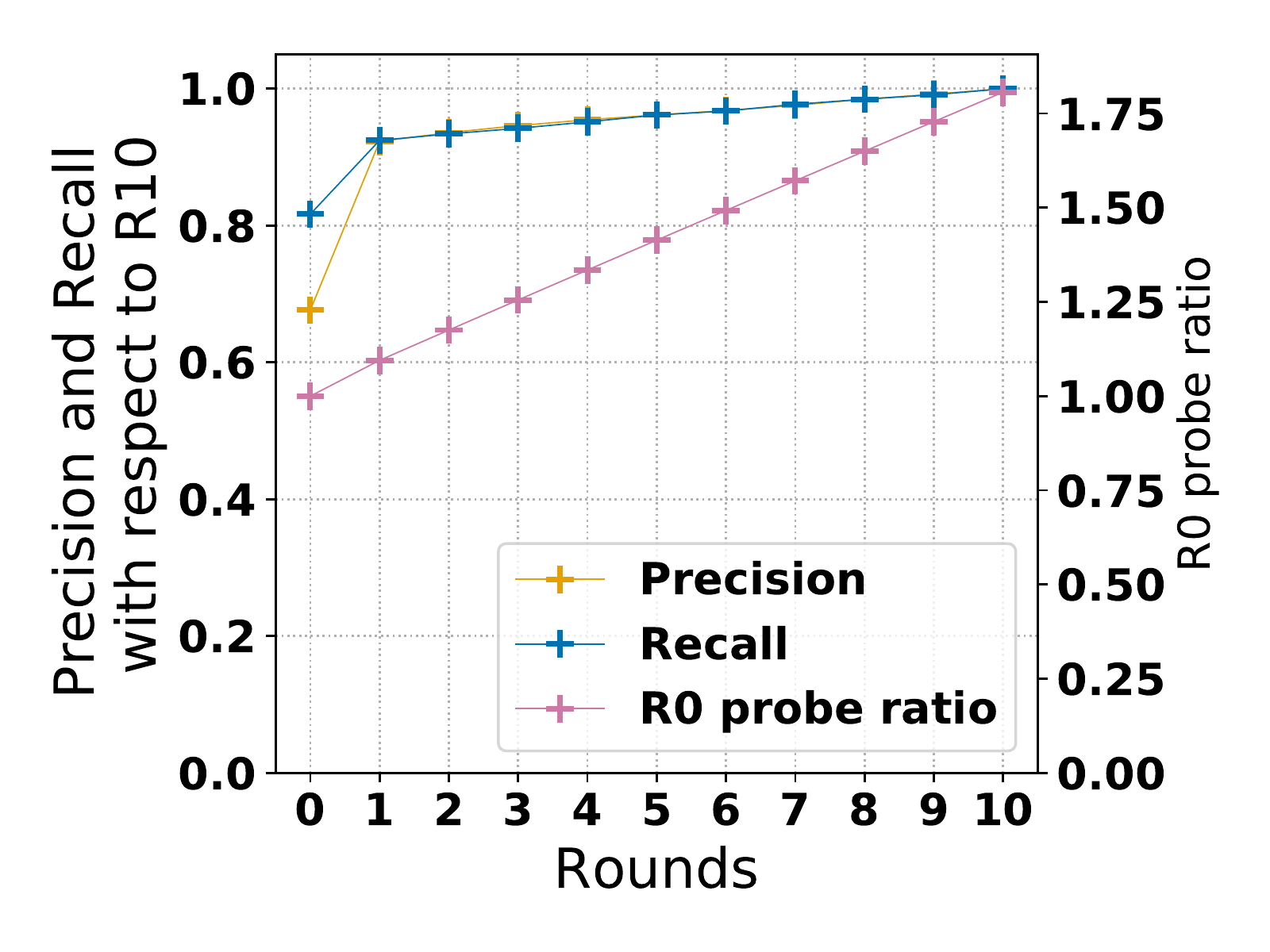}}
  \end{subfigure}
  \caption{Alias resolution over ten rounds}
  \label{fig:alias-resolution-additional}
  \vspace{-2em}
\end{figure}

We looked at how MMLPT's alias resolution results evolve round by
round. Round 0 is based on just the data obtained through MDA-Lite
Paris Traceroute, with no additional probing. The MBT and
signature-based tests are applied to the extent possible. Round 1 adds
one direct probe to each of the IP addresses at a given hop, in order
to provide more complete Network Fingerprinting signatures. It
also is the first round of MBT probing,
attempting to elicit 30 replies per address. Each
subsequent round through to Round 10 consists of an additional 30
indirect probes per address, in order to further refine the alias sets
using the MBT.

Fig.~\ref{fig:alias-resolution-additional} presents overall values
for precision, recall, and numbers of probes sent over the
10,000 measurements conducted for the MDA-Lite evaluation of
Sec.~\ref{mda-lite-evaluation}. We do not have ground truth, so
precision and recall are relative to our best available determination
of the alias sets, which is the result of Round 10 in each case. The
number of probes is relative to the number sent in Round 0.

Round 0, with no probing beyond that which is performed for MDA-Lite
Paris Traceroute, yielded 68\% precision and 81\% recall with respect
to the Round 10 results. A significant jump to 92\% in both cases came
with a first round of probing, and then there was a slow increase with
each successive round. The additional probing for each round was less
than 10\% of the basic MDA-Lite Paris Traceroute probing. 

These results indicate that we can glean router-level information with
a modest amount of additional probing, typically 20\% more is enough to get 
a precision and a recall greater than 92\% with respect to round 10, 
and 75\% more to complete
the ten rounds.
Additional work will be
required in order to better establish a firm basis against which to
compare, so as to provide clearer guidance on the tradeoff between
probing and the completeness and accuracy of the results.

{\setlength{\extrarowheight}{5pt}
\begin{table}
\begin{center}
\resizebox{\linewidth}{!}{%
\begin{tabular}{|l|r|r|r|}
\cline{2-4}
    \multicolumn{1}{l|}{} & Accept Direct & Reject Direct & Unable Direct\\
\hline
        Accept Indirect & 0.365 & 0.005 & 0.283\\
\hline
	Reject Indirect & 0.144 & N/A & N/A\\
\hline
	Unable Indirect & 0.203 & N/A & N/A\\
\hline
\end{tabular}}
\caption{Findings for 4798 address sets identified as routers either
  by indirect probing (MMLPT) or direct probing (\textsc{Midar}), expressed as
  portions adding up to 1.0}
\label{table:midar-comparison}
\end{center}
\vspace{-3em}
\end{table}}

We also looked at the potential benefits and costs of adding direct
probing, as we had implemented MMLPT with only indirect probing, for
the MBT. For each diamond, MMLPT identifies zero or more address sets
as routers, validating or rejecting address sets via indirect probing.
We compare these results with what direct probing IP ID techniques
would have found, using \textsc{Midar} for this. We ran \textsc{Midar}
on all the addresses of the diamond, and \textsc{Midar} too identifies
zero or more address sets as routers in the diamond. We take the union
of the address sets identified by both tools, and compare: which ones
did both accept as being a router, and which ones were accepted by one
of the tools but not by the other? If a tool does not accept an
address set, it is either because it has rejected it (for instance by
finding a pair of addresses that has failed the MBT) or because it is
unable to determine if one or more of the addresses belongs in the set
(for instance because of an insufficient time series from an address).

Table~\ref{table:midar-comparison} shows the results for 4798 address
sets, of which 3414 were identified as routers by \textsc{Midar} and
3140 by MML\-PT. The values are the portion of address sets that fall
within each category. 36.5\% were accepted as routers by both
\textsc{Midar} and MMLPT. Just 0.5\% of sets accepted by MMLPT are
rejected by \textsc{Midar}, whereas 14.4\% of sets accepted by
\textsc{Midar} are rejected by MMLPT. The latter can be explained by
routers that implement per-interface counters for the IP ID for the
ICMP Time Exceeded messages associated with indirect probing and
router-wide counters for the ICMP Echo Reply messages associated with
direct probing. 

Significant portions of sets accepted by one tool encounter a failure
to determine a result by the other tool: 20.3\% of sets accepted by
\textsc{Midar} led to no conclusion by MMLPT and 28.3\% of sets
accepted by MMLPT led to no conclusion by \textsc{Midar}. Upon further
investigation, we found that 98.6\% of the non conclusive cases for
MMLPT are due to either constant (mostly zero) IP IDs and 1.4\% to non
monotonic IP ID series. Looking at \textsc{Midar} logs, we found that
the 28.3\% inconclusive cases had different causes: for each
inconclusive set, at least one IP in the set was either unresponsive
to direct probing (60.5\%), or its IP ID series was a copy of the
probe IP ID (22.8\%), or its IP ID series was non monotonic (13.6\%),
or \textsc{Midar} got unexpected responses, meaning that the reply did
not match that which would be expected based upon the probe protocol used (3.1\%).

Our overall conclusion is that direct probing provides a potentially
valuable complement to indirect probing, and that we should include it
in future versions of MMLPT, while also evaluating the tradeoff in
what is gained against the additional probing cost that it will
entail.

\section{Surveys}\label{surveys}\label{metrics}

This section presents the two surveys that we have conducted, one at
the IP level, the other at the router level. The aim in both is to
characterize the topologies that are encountered by multipath route
tracing in the IPv4 Internet, along the lines of earlier
surveys~\cite{Augustin:2011:MMR:2042972.2042989,marchetta2016and,almeida2017characterization}
mentioned in the Related Work section.

Our focus is on the ``diamonds'' (see Sec.~\ref{mda-recap} for the
definition) that are encountered in a route trace. We define a a
\textbf{distinct diamond} by its divergence point and its convergence
point. This means that if a diamond is encountered multiple times in
the course of a survey, there might be differences in its measured
internal topology from one encounter to the next. If either a
divergence point or a convergence point is non-responsive (a ``star''
in common parlance), we consider it as different from a diamond that
has responsive divergence and convergence points, even if the two
diamonds have other IP addresses in common. Since a diamond might show
up in multiple measurements, we define each encounter with a distinct
diamond to be a \textbf{measured diamond}. Each way of counting
reflects a different view of what is important to consider: the number
of such topologies, or the likelihood of encountering one. We look at
both.

The surveys describe how large diamonds are, both in number of hops
and in number of vertices at a given hop. Also, because we have found
that ``uniformity'' and ``meshing'' are relevant to the ability to
economize on probes when tracing at the IP level (see
Sec.~\ref{uniformity-and-meshing}), we describe these features. For
the metric definitions that follow, we apply those of Augustin
\etal~\cite{Augustin:2011:MMR:2042972.2042989} for ``maximum width''
and ``maximum length'' and add ``maximum width asymmetry'' and ``ratio
of meshed hops''. 
As illustrated in Fig.~\ref{fig:diamond-metrics}, these are:
\begin{figure}[b]
\vspace{-3mm}
\begin{center}
	\includegraphics[width=1\columnwidth]{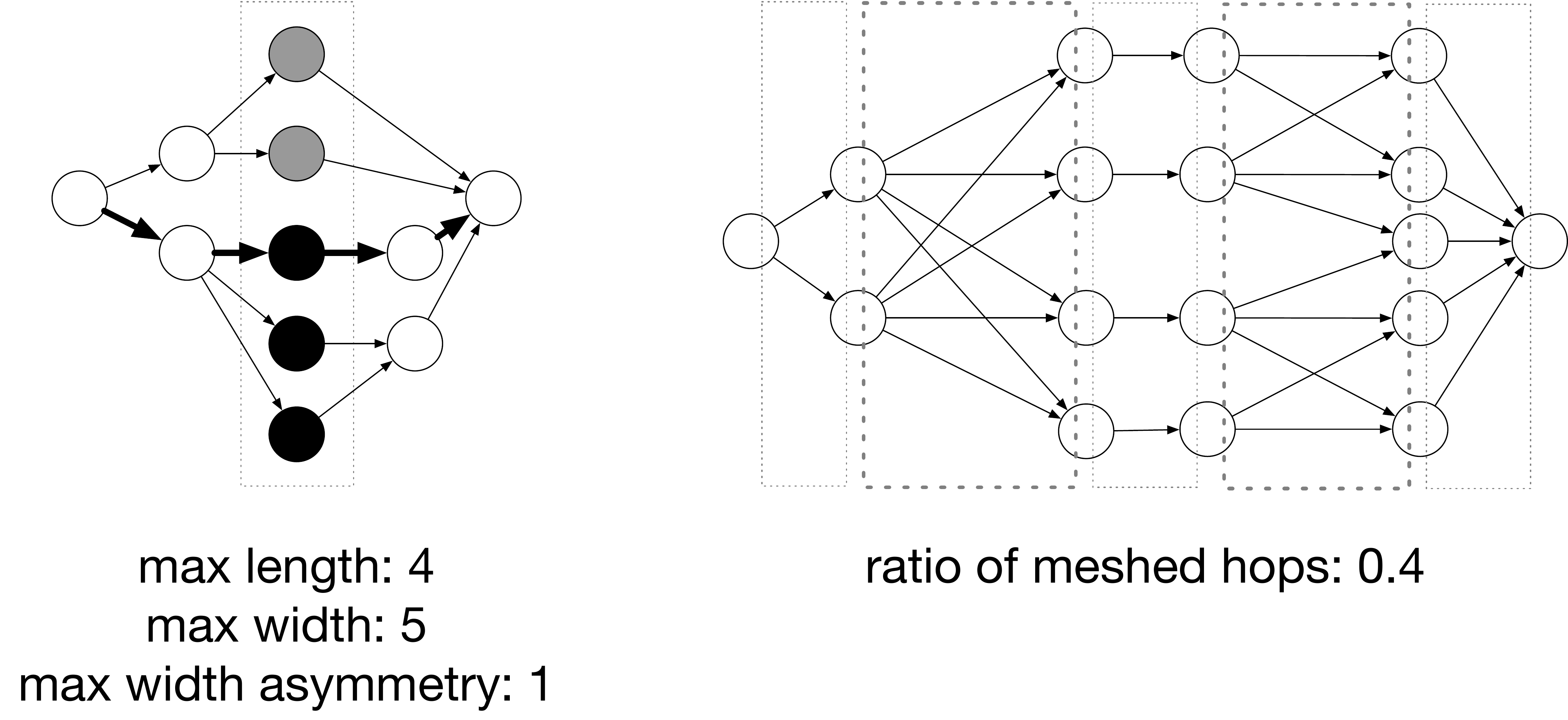}
	\caption{Diamond metrics}
	\label{fig:diamond-metrics}
\end{center}
\end{figure}

The \textbf{maximum width} is the maximum number of vertices that can
be found at a single hop, as in the boxed hop of the left-hand diamond.

The \textbf{maximum length} is the length of the longest path between
the divergence and the convergence point, as shown by the set of bold
edges in the left-hand diamond.

The \textbf{maximum width asymmetry} is a topological indicator of a
lack of uniformity. We define it first for a pair of hops~$i$ and $i+1$.
\begin{itemize}
\item If hop~$i$ has fewer vertices than hop~$i+1$, it is the maximum
  difference in the number of successors between two vertices at hop
  $i$.
\item If hop~$i$ has more vertices than hop~$i+1$, it is the maximum
  difference in the number of predecessors between two vertices at hop
  $i+1$.
\item If hops~$i$ and $i+1$ have identical numbers of vertices, it is
  the maximum of the two values described above.
\end{itemize}
For a diamond as a whole, it is the largest value of maximum width
asymmetry found across all hop pairs, as shown by the grey
and black vertices of the left-hand diamond.

The \textbf{ratio of meshed hops} of a diamond is the portion of hop
pairs of hops that are meshed, as shown in the right-hand diamond, in
which two of the five hop pairs are meshed, for a ratio of 0.4.

\subsection{IP level survey}\label{mda-survey}

The IP level survey is based on multipath route traces from 35 sources
towards 350,000 destinations during two weeks starting 8 March 2018.

The route tracing tool was the \textit{libparistraceroute}-based MDA
Paris Traceroute~\cite{libparistraceroute}, using its default
parameters. We employed UDP probes, as Luckie
\etal~\cite{Luckie:2008:TPM:1452520.1452557} found best results for
discovering load balanced paths with such probes.

The sources were PlanetLab nodes running Fedora~24 or 25, obtained
through PlanetLab Europe~\cite{ple}. (We also ran a survey with similar results,
which can be found at the URL mentioned at the end of Sec.~\ref{intro}, on the
new EdgeNet infrastructure~\cite{edgenet} affiliated with PlanetLab Europe.)

The destinations were chosen at random from the IPv4 addresses rated
as ``highly responsive'' in the Internet Address Hitlist \textsc{Impact}
dataset {\small\texttt{Internet\_address\_hitlist\_\-it78w-20171113}},
ID DS-822, covering 17 January 2015 to 15~December 2017~\cite{usc-dataset}.

We discarded route traces that we could not collect because of 
infrastructure troubles, 
yielding
294,832 exploitable results, among which 155,030 passed through at
least one per-flow load balancer. There were 60,921 distinct and
220,193 measured diamonds.

\begin{figure}[h]
\begin{subfigure}{.25\textwidth}
  \centering
  \includegraphics[width=\linewidth]{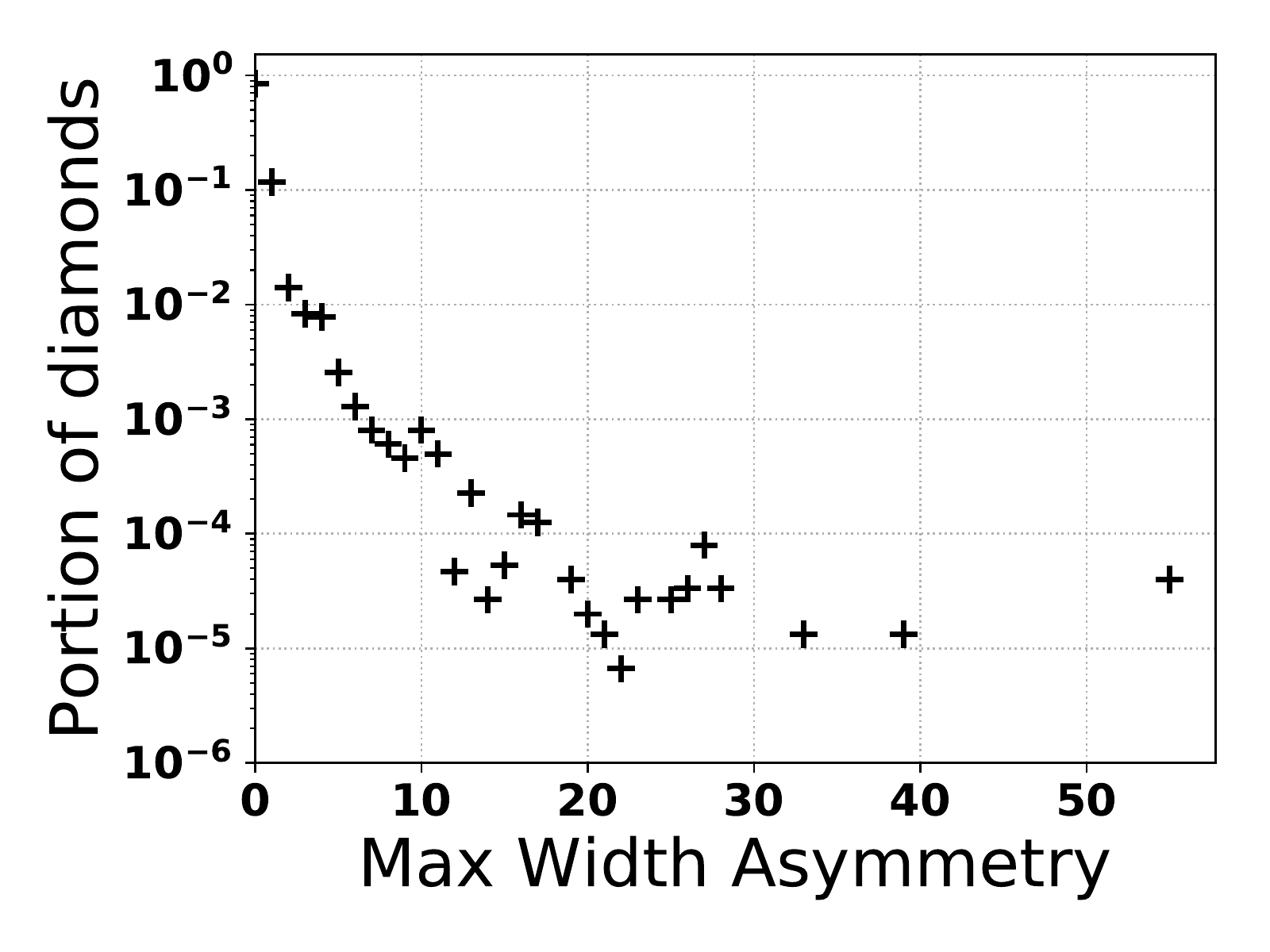}
	\caption{Measured}  
\end{subfigure}%
\begin{subfigure}{.25\textwidth}
  \centering
  \includegraphics[width=\linewidth]{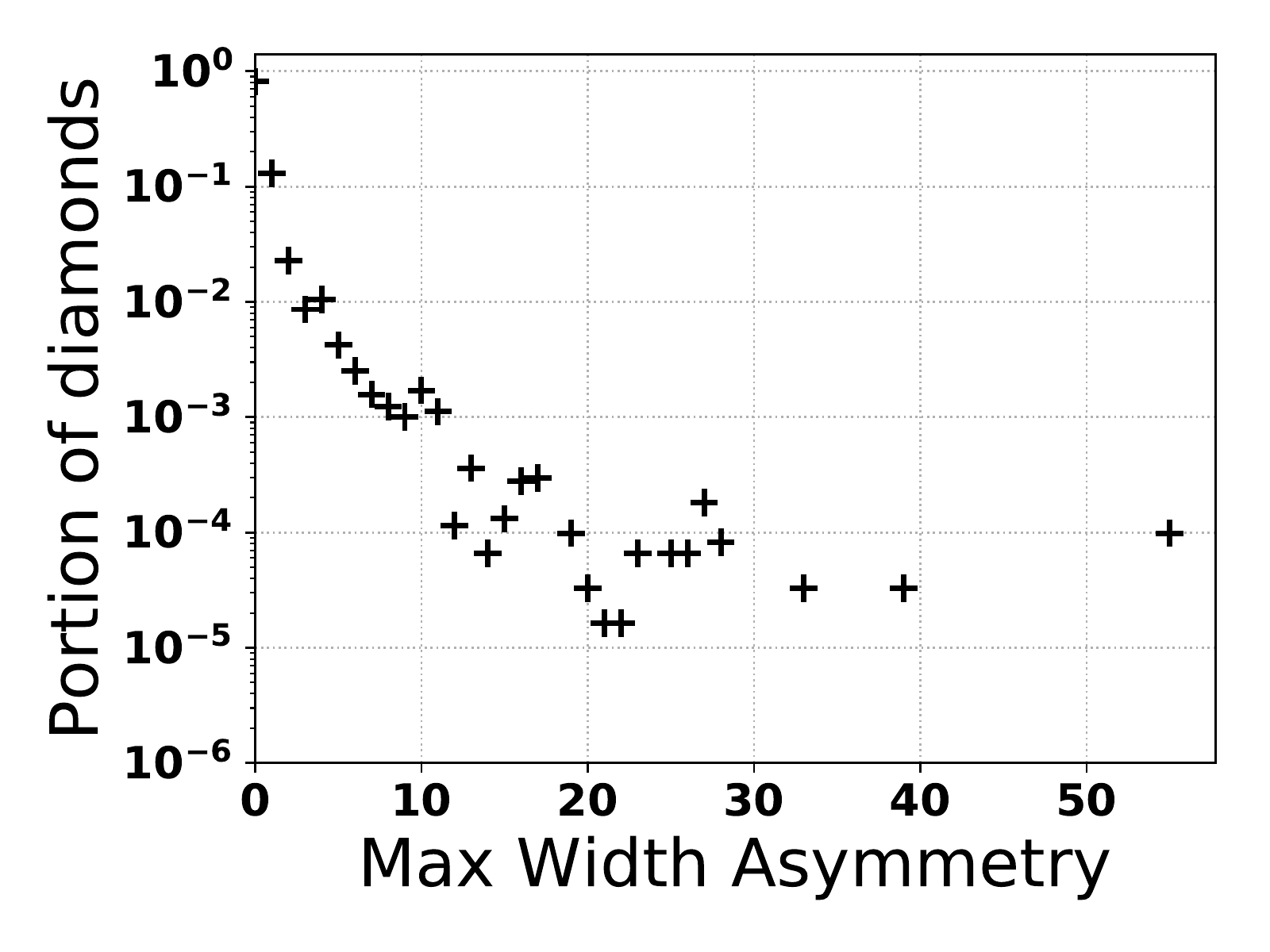}
  	\caption{Distinct}  
\end{subfigure}
\caption{Width asymmetry}
\label{fig:max-width-asymmetry}
\end{figure}

\begin{figure}[h]
\begin{subfigure}{.25\textwidth}
  \centering
  \includegraphics[width=\linewidth]{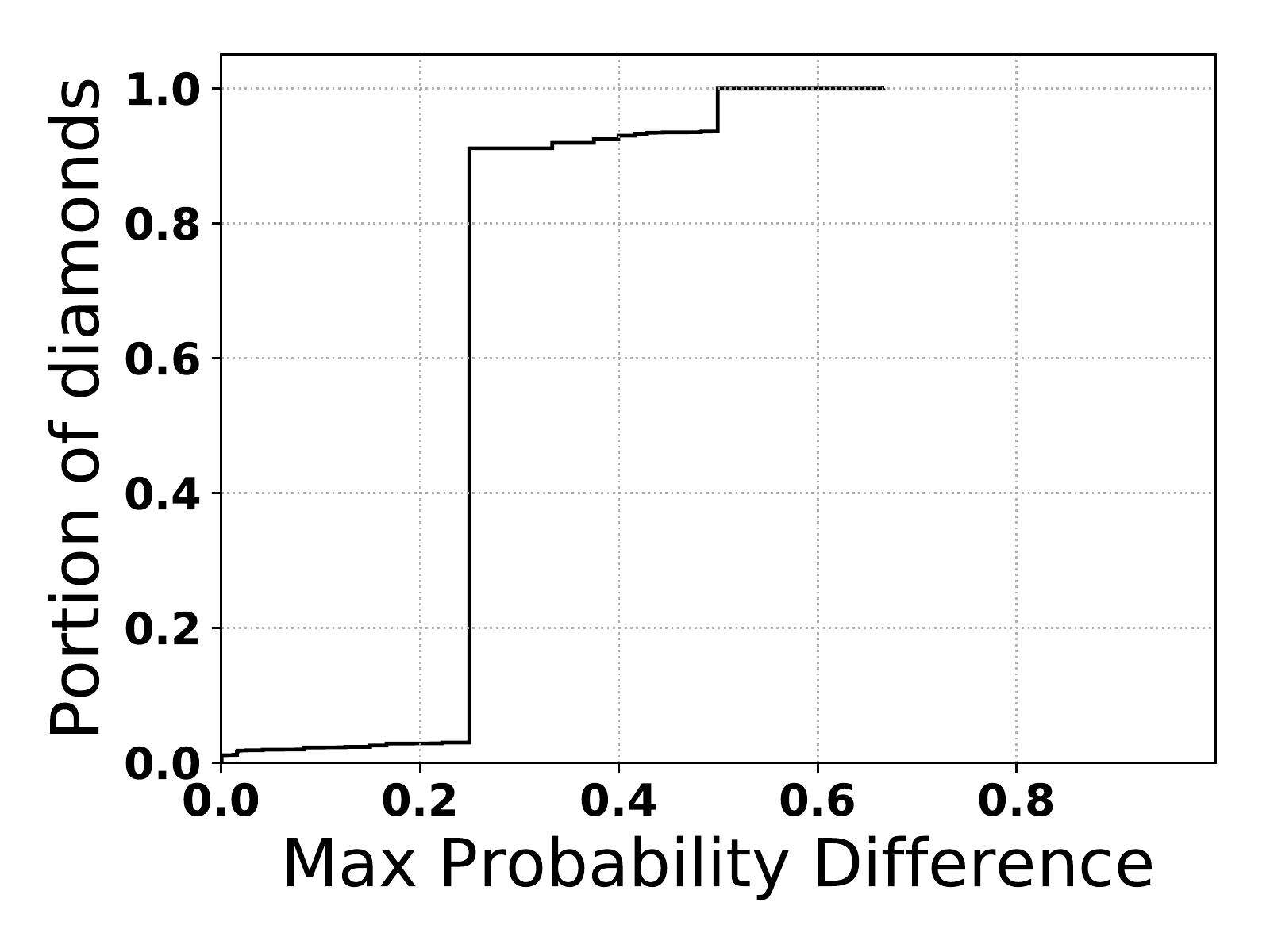}
	\caption{Measured}  
\end{subfigure}%
\begin{subfigure}{.25\textwidth}
  \centering
  \includegraphics[width=\linewidth]{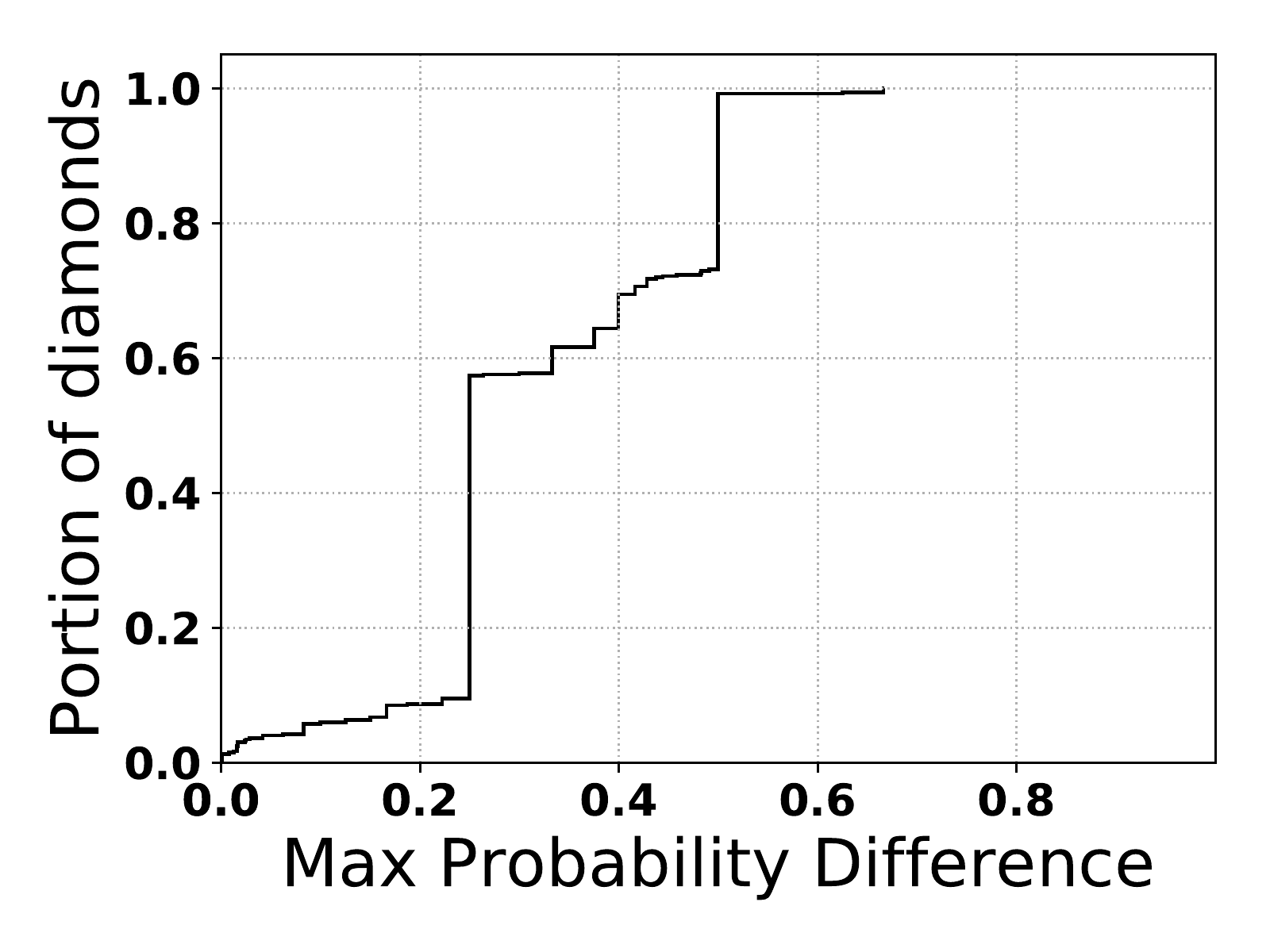}
	\caption{Distinct}  
\end{subfigure}
\caption{Maximum probability difference in width-asymmetric diamonds}
\label{fig:max-probability-difference}
\end{figure}

We start by looking at uniformity and meshing.

\paragraph{Uniformity}\label{important-results}
In both measured and distinct diamond asymmetry distributions
(Fig.~\ref{fig:max-width-asymmetry}), 89\% of diamonds have zero
asymmetry. This means that most diamonds are uniform, provided that
load balancing is uniform across next hop interfaces, and supports the
MDA-Lite's assumption of uniformity. But if the MDA-Lite cannot detect
the asymmetry in a diamond that is among the 11\% that are asymmetric,
it will not switch over to the full MDA and it risks failing to
discover the full topology. It is most likely to encounter difficulty
on an unmeshed diamond, as, when meshing is detected, the full MDA is
invoked. Only 2.3\% of measured and 3.6\% of distinct diamonds are
both asymmetric and unmeshed. We examined these diamonds for
differences in discovery probability among vertices at a common hop,
plotting the CDFs of all non-zero probability differences in
Fig.~\ref{fig:max-probability-difference}. In these cases, 90\% of
measured and 58\% of distinct diamonds have a maximum probability
difference of 0.25 and, for both, 99\% have a maximum probability
difference of 0.5. This indicates that the MDA-Lite is very unlikely to
fail in uncovering a lack of uniformity, which is borne out by our
experimental results in Sec.~\ref{mda-lite-evaluation}. This issue
could be more rigorously studied with further mathematical analysis.

\paragraph{Meshing}\label{meshing-results}
\begin{figure}[h]
\begin{subfigure}{.24\textwidth}
  \centering
  \includegraphics[width=\linewidth]{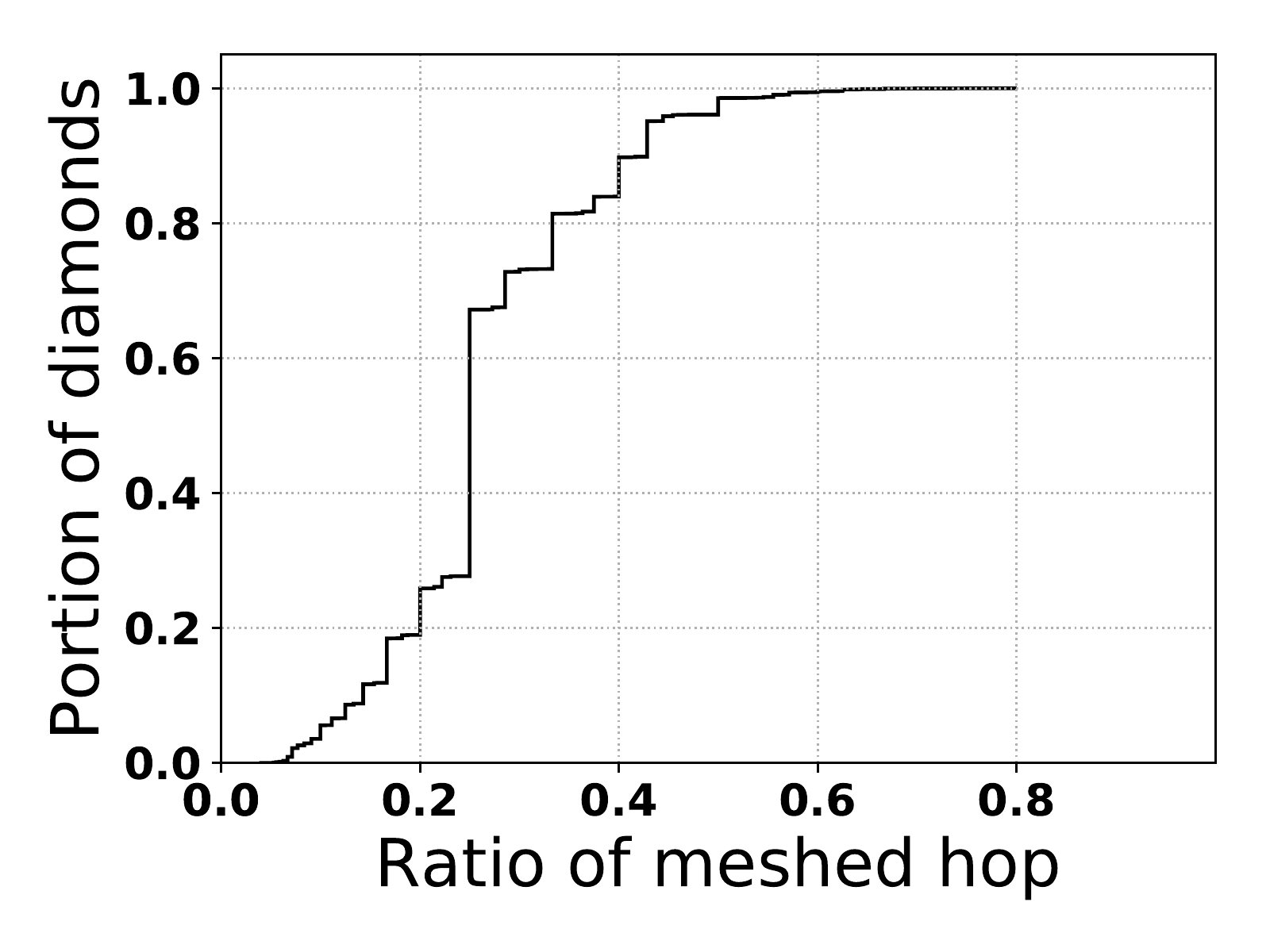}
	\caption{Measured}  
\end{subfigure}%
\begin{subfigure}{.24\textwidth}
  \centering
  \includegraphics[width=\linewidth]{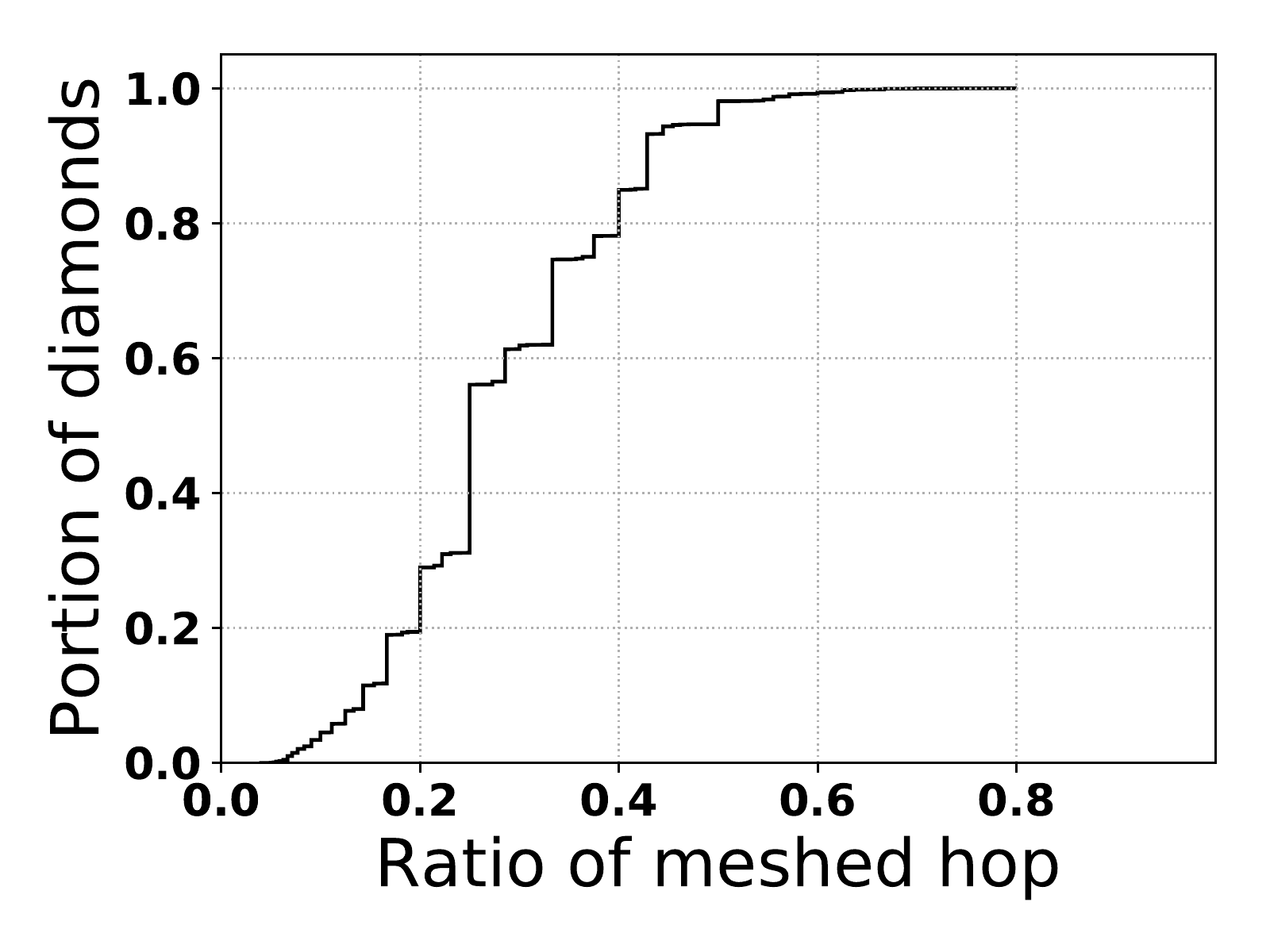}
	\caption{Distinct}  
\end{subfigure}
\caption{Ratio of meshed hops}
\label{fig:meshed-hops}
\end{figure}

\begin{figure}[h]
\begin{subfigure}{.24\textwidth}
  \centering
  \includegraphics[width=\linewidth]{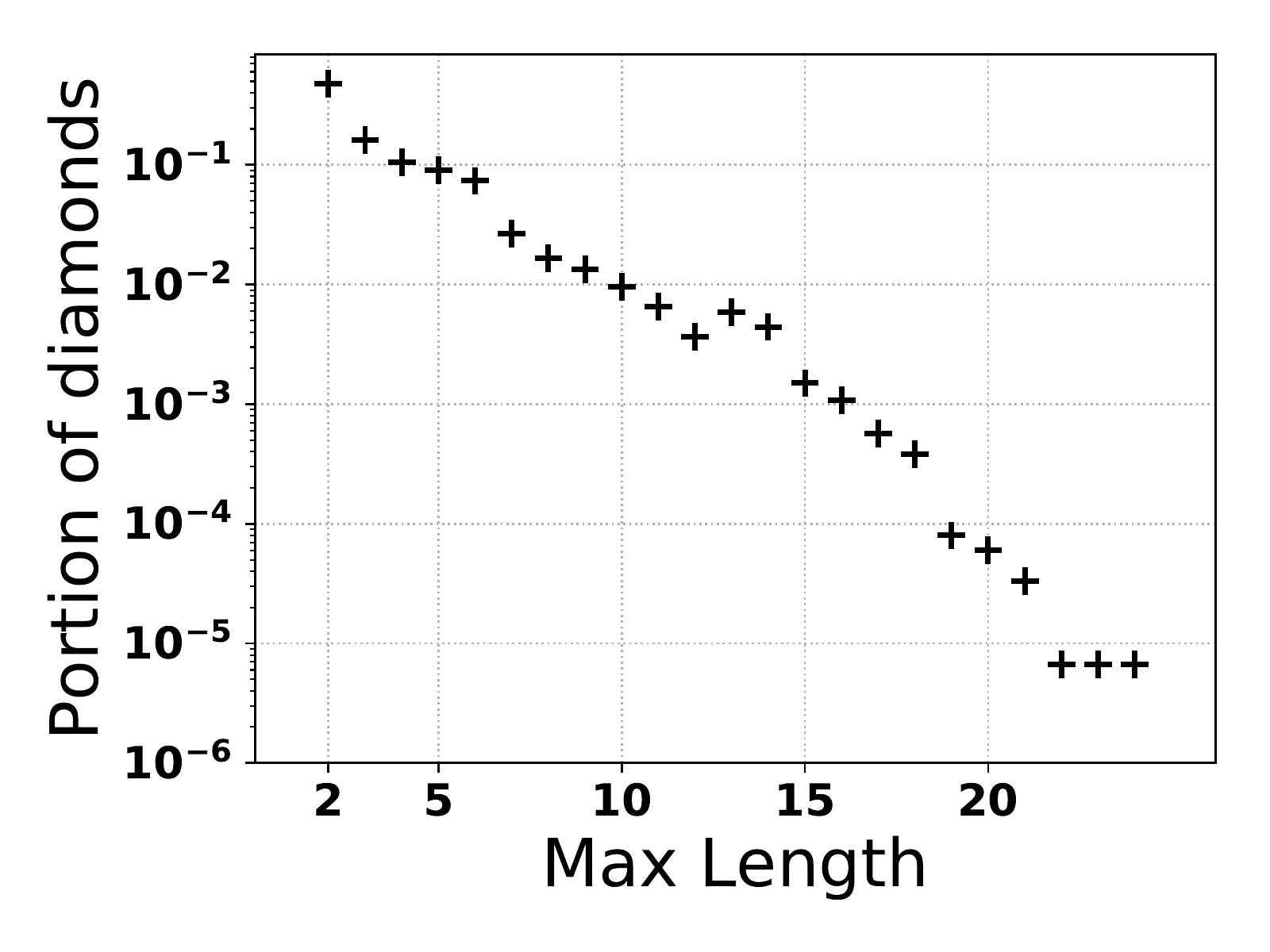}
\end{subfigure}%
\begin{subfigure}{.24\textwidth}
  \centering
  \includegraphics[width=\linewidth]{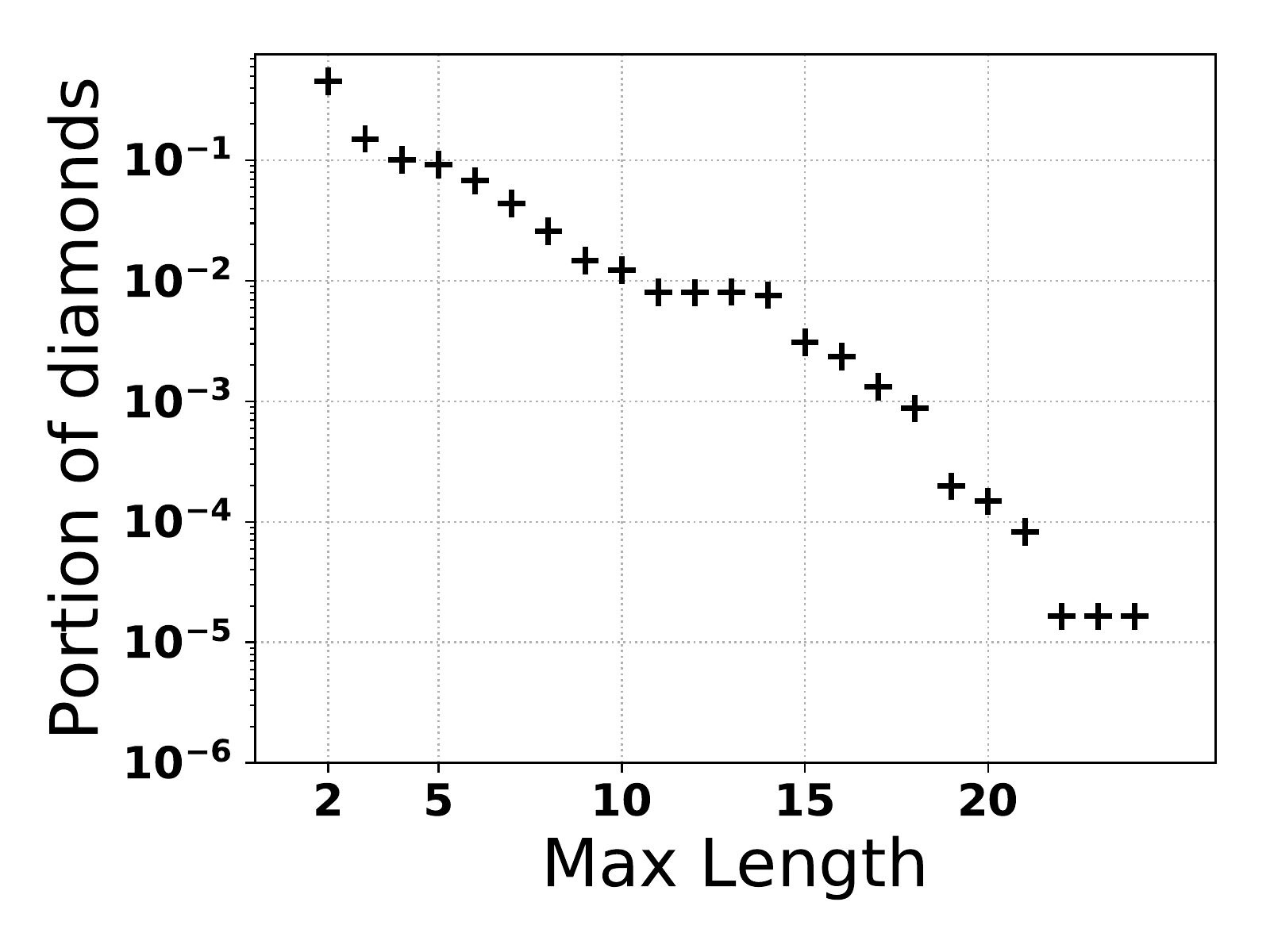}
\end{subfigure}

\begin{subfigure}{.24\textwidth}
  \centering
  \includegraphics[width=\linewidth]{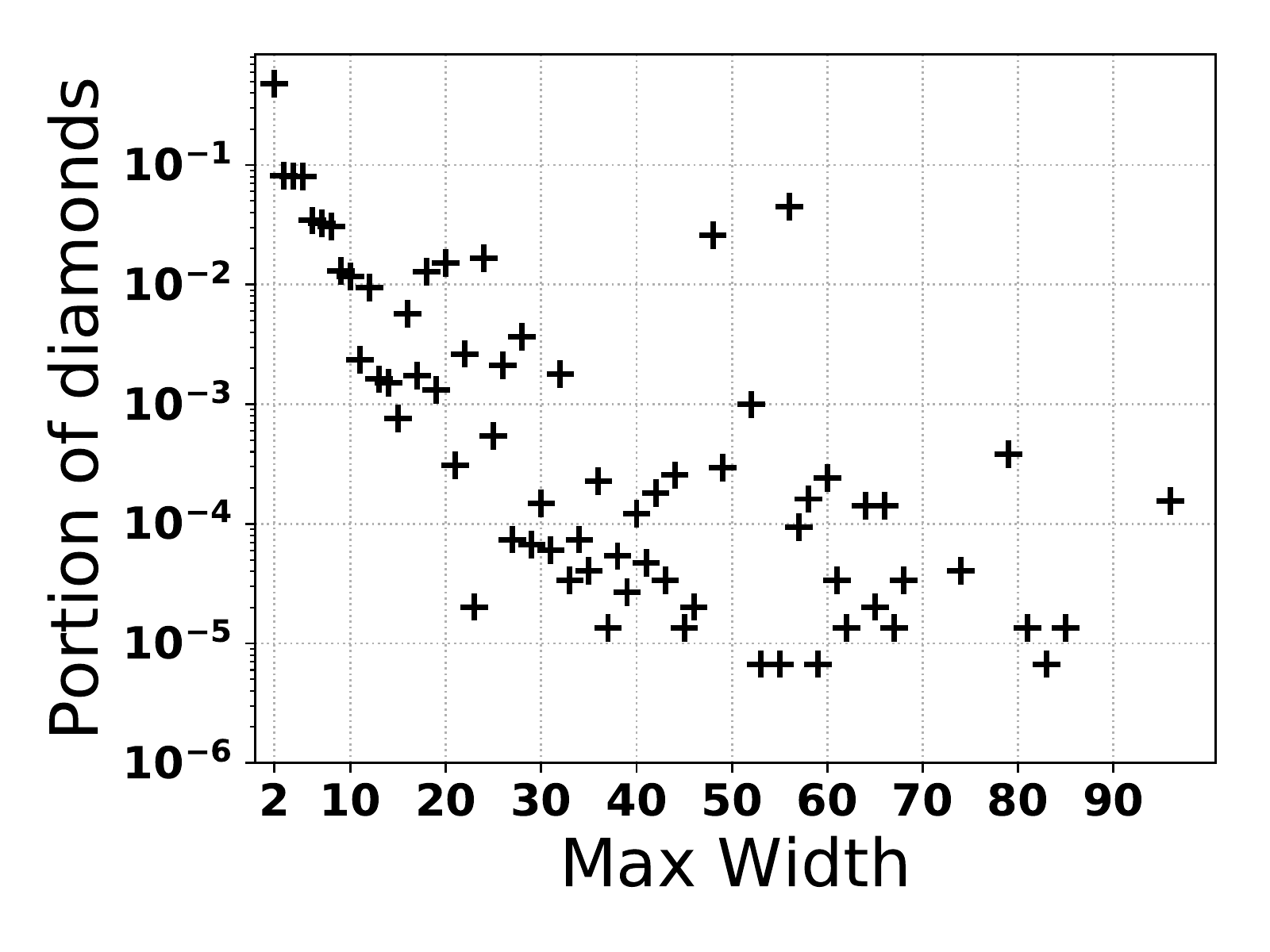}
	\caption{Measured}  
\end{subfigure}%
\begin{subfigure}{.24\textwidth}
  \centering
  \includegraphics[width=\linewidth]{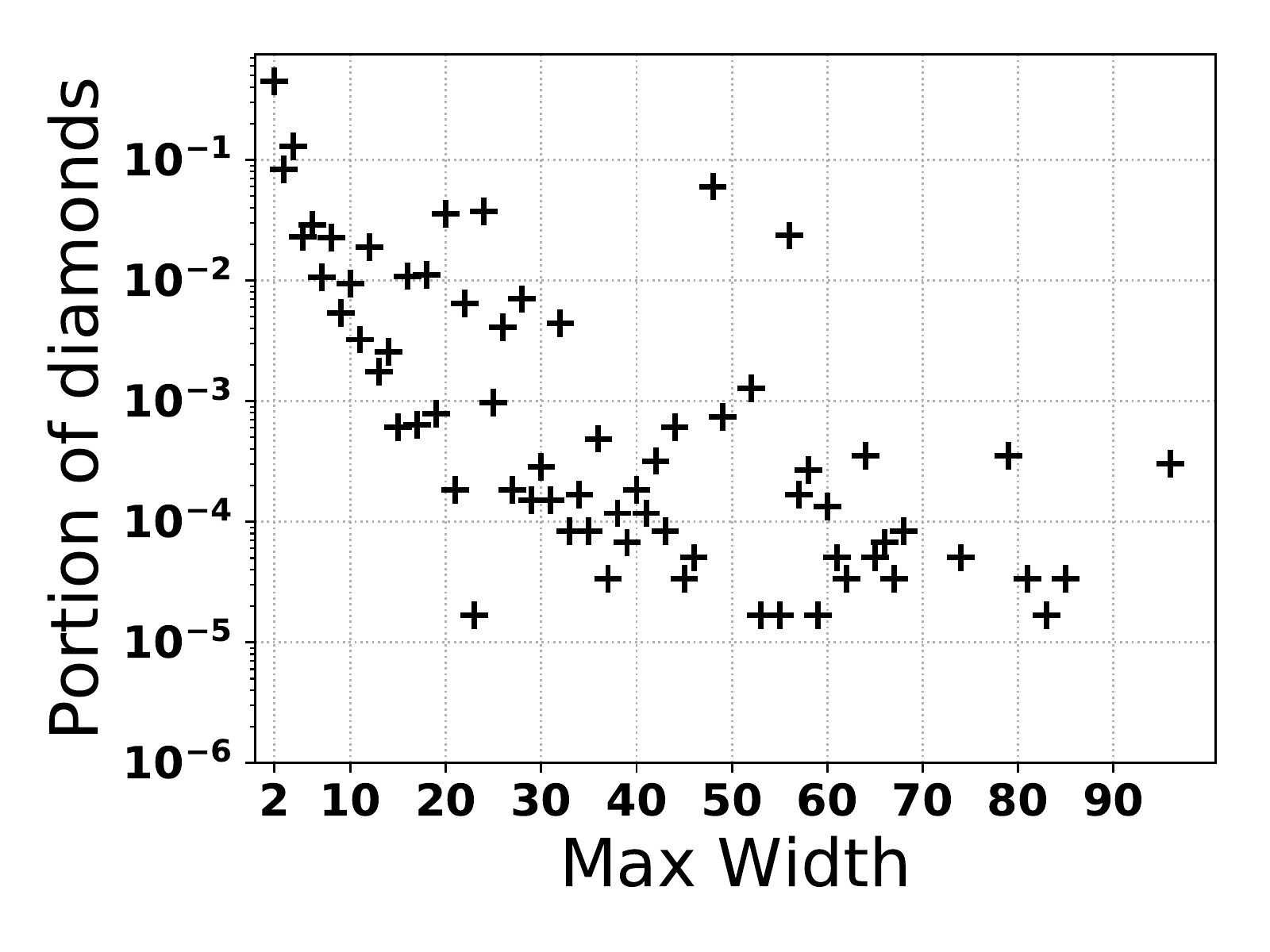}
	\caption{Distinct}  
\end{subfigure}
\caption{Maximum length and maximum width}
\label{fig:metrics-distribution}
\end{figure}

Of the 220,193 measured diamonds in our survey, 32,430 present at
least one meshed hop, and of the 60,921 distinct diamonds, 19,138 are
meshed. Fig.~\ref{fig:meshed-hops} plots CDFs of the ratio of meshed
hops for the meshed diamonds. The MDA-Lite offers probe savings over
the full MDA when a pair of hops is not meshed. More than 80\% of
meshed diamonds have a ratio of of meshed hops under 0.4, which
indicates a significant potential for the MDA-Lite to realize
significant probe savings, even on meshed diamonds.
We continue by looking at the length and width metrics, for which the
distributions are shown in Fig.~\ref{fig:metrics-distribution}.
Almost half of both measured and distinct diamonds have a maximum
length of 2, meaning that they consist of a divergence point, a single
multi-vertex hop, and a convergence point. The MDA-Lite is more
economical than the full MDA on such diamonds.
The largest value of maximum width encountered is 96. Such a high
value is unprecedented, with earlier
surveys~\cite{Augustin:2011:MMR:2042972.2042989,marchetta2016and}
reporting maximum widths of at most 16.
A notable feature of the maximum width distributions is their peaks at
48 and 56. Further investigation indicates that the distinct diamond
distribution might be overstating what is in fact being encountered by
the route traces. Though the diamonds are distinct by our definition,
meaning that they have a unique pair of divergence and convergence
points, they share a large portion of their IP addresses. This
suggests a common structure that is being frequently encountered
via a variety of ingress points.

\begin{figure}[h]
\begin{subfigure}{.24\textwidth}
  \centering
  \includegraphics[width=\linewidth]{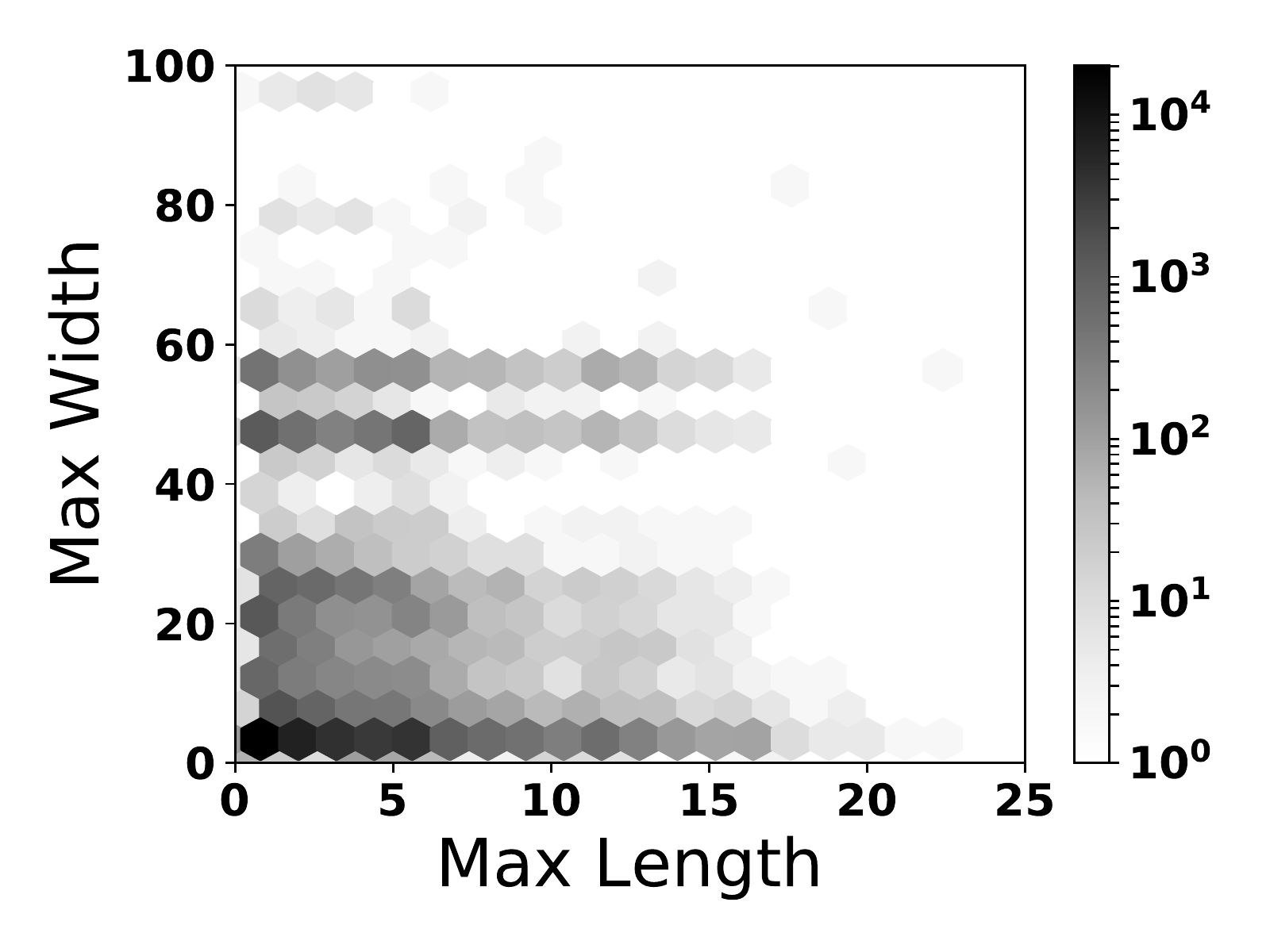}
	\caption{Measured}  
\end{subfigure}%
\begin{subfigure}{.24\textwidth}
  \centering
  \includegraphics[width=\linewidth]{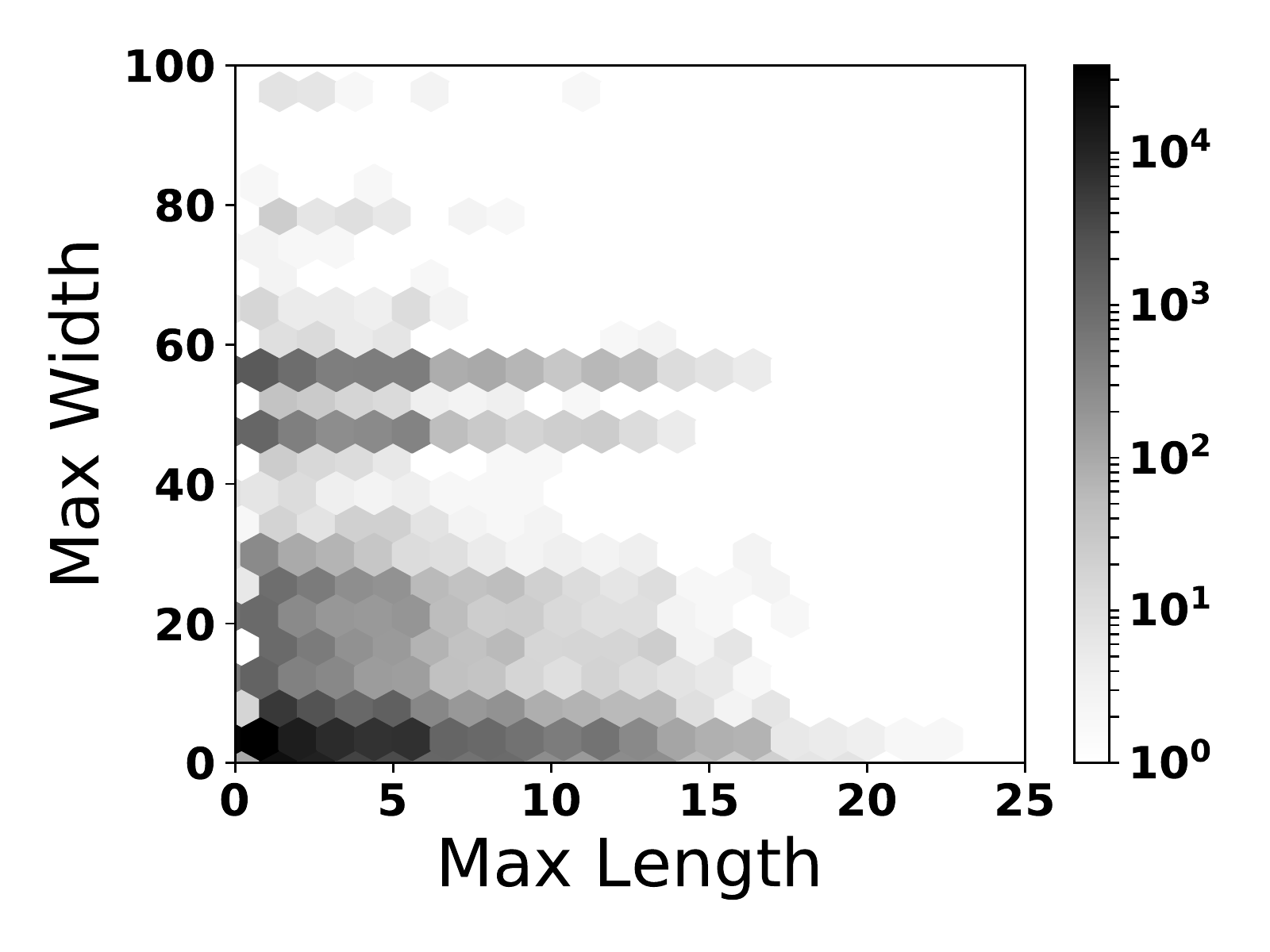}
	\caption{Distinct}  
\end{subfigure}
\caption{Maximum length and maximum width joint distributions}
\label{fig:max-length-max-width}
\end{figure}
Looking at the joint distributions of maximum width and maximum
length (Fig.~\ref{fig:max-length-max-width}), we see that short and
narrow diamonds continue to be the most common, as found in previous
surveys. For example, we found that 24.2\% of measured and 27.4\% of
distinct diamonds were of maximum length 2 and maximum width 2,
corresponding to the simplest possible diamond. The maximum width 48
and 56 diamonds also reveal themselves to have a variety of different
maximum lengths.  

\subsection{Router level survey}\label{router-survey}
The router level survey is based upon the 155,030 route traces from
the IP level survey that passed through at least one load balancer. We
retraced these with Mutilevel MDA-Lite Paris Traceroute during two
weeks, starting on 3 April 2018. For each trace, we obtained IP level
output and router level output.

\begin{figure}[h]
\begin{subfigure}{.25\textwidth}
  \centering
  \includegraphics[width=\linewidth]{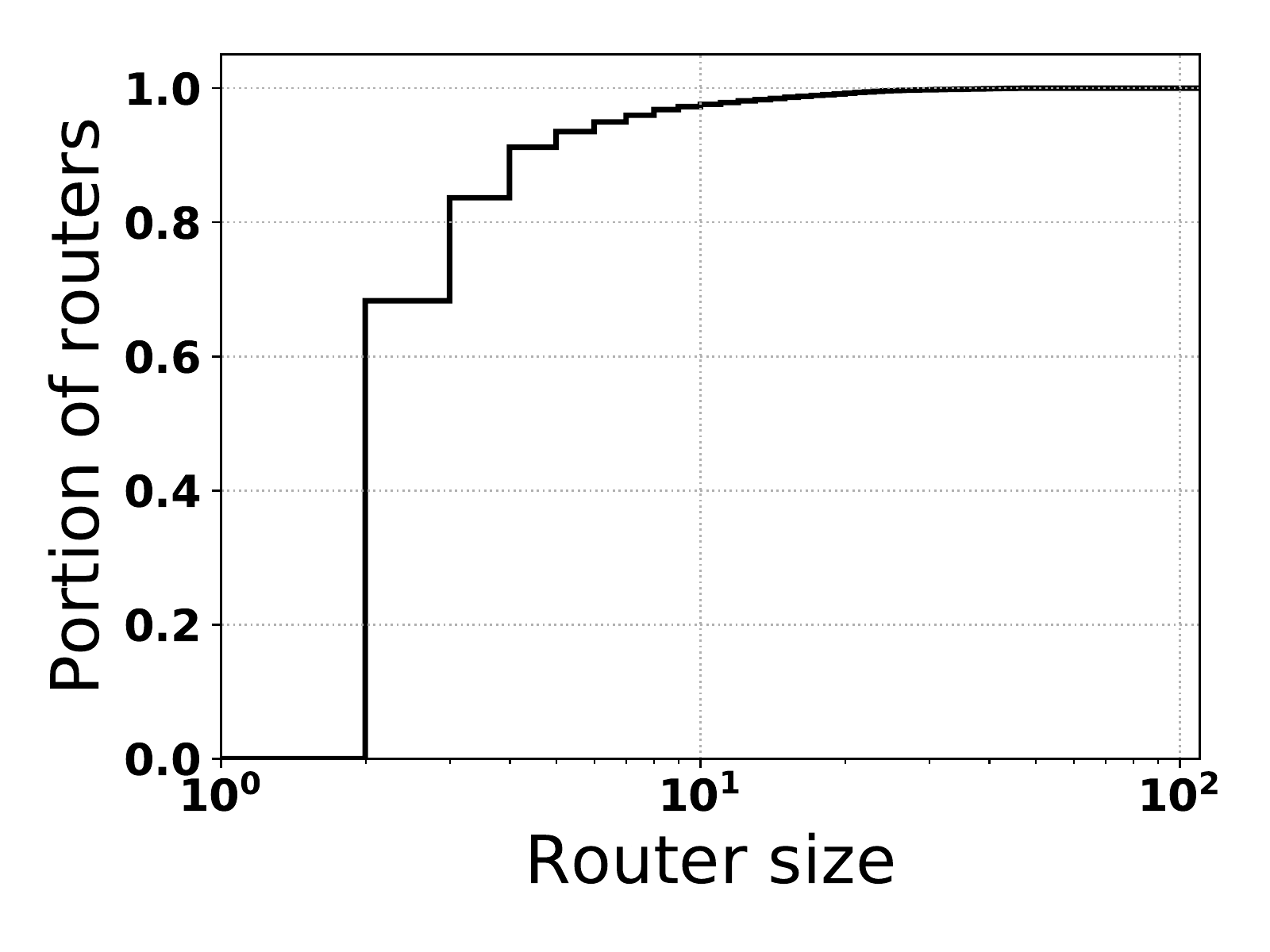}
  \caption{Distinct routers}
\end{subfigure}%
\begin{subfigure}{.25\textwidth}
  \centering
  \includegraphics[width=\linewidth]{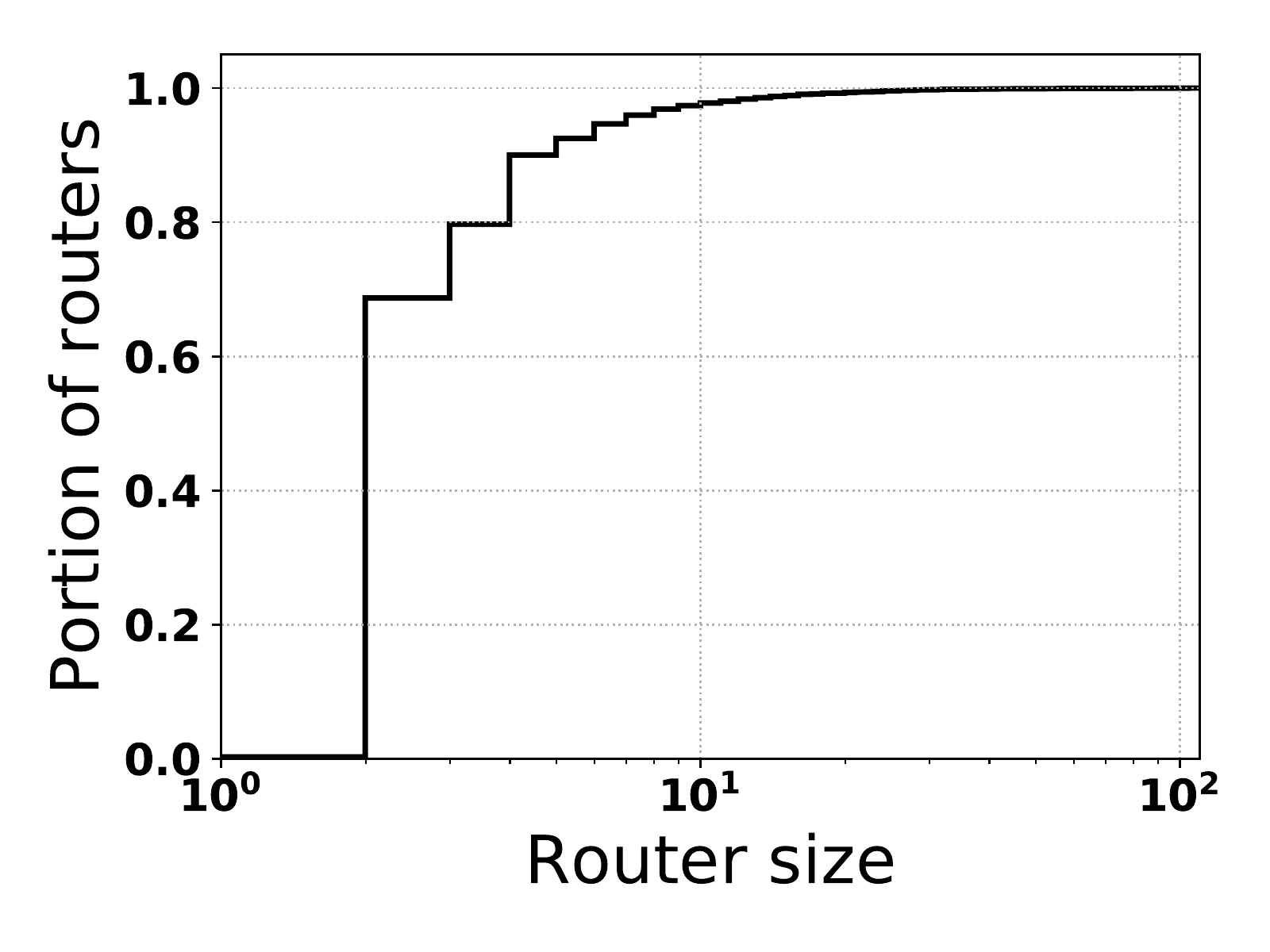}
  \caption{Aggregated routers}
\end{subfigure}
\caption{Router size}
\label{fig:router-size-cdf}
\end{figure}
We found 646 cases of distinct address sets (0.98\% of
the total alias set) that were considered as aliases 
by one measurement, but
discarded or not found by another, although they had both seen
the entire address set at the IP level. A deeper analysis showed that 295 of
those cases
were due
to a constant 0 IP ID series collected by one measurement for at least one address
in the address set, 
whereas the other measurement could build a monotonic IP ID time series
for each of the addresses in the address set. The 
remaining 351 cases were false positives, which were then discarded from the router
dataset analysed in this section.

We looked at what we term the ``size'' of the routers that were found,
the size being the number of IP interfaces identified as belonging to
a router. A route trace from a given vantage point is bound to pick up
mostly the ingress interfaces facing that point, which tend to be the
ones from which it receives responses,
and so this metric will be an underestimate of
the true number of interfaces. We also aggregated the IP interface
sets from multiple traces through transitive closure based upon two
sets having at least one address in common, which may give less of an
underestimate, but is still incomplete, as we do not perform
full alias resolution on the overall IP addresses set found. 
CDFs of the sizes are shown in
Fig.~\ref{fig:router-size-cdf}.  68\% of the routers had a size of 2
and 97\% had a size of 10 or less.  We found 1 distinct router with
more than 50 interfaces, and 5 such routers when we aggregated the
address sets.

We looked at what happens to each IP level diamond when it is resolved
into a router level diamond. There are four possibilities: (1) there
is no alias resolution, so the diamond remains the same; (2) the
diamond resolves into a single smaller diamond; (3) the diamond
resolves into a series of smaller diamonds; (4) the diamond disappears
completely, being resolved into a straight path of routers.  As
Table~\ref{table:collapse-routers} shows, some degree of router
resolution takes place on 41.9\% of unique diamonds. In comparison,
Marchetta~\etal~\cite{marchetta2016and} saw, in 2016, a 33\% reduction
in diamond max-width, when applying \textsc{Midar} a posteriori to multipath
route traces.

{\setlength{\extrarowheight}{5pt}
\begin{table}
\resizebox{0.6\linewidth}{!}{%
\begin{tabular}{|l|l|}
\hline
    \thead{Case} & \thead{Fraction}\\
\hline
No change & 0.579 \\      
\hline
Single smaller diamond &  0.355\\
\hline
Multiple smaller diamonds & 0.006\\
\hline
One path (no diamond) & 0.058\\
\hline
\end{tabular}}
\vspace{1em}
\caption{Effect of alias resolution on unique diamonds}
\label{table:collapse-routers}
\vspace{-1em}
\end{table}}

\begin{figure}[h]
\begin{subfigure}{.25\textwidth}
  \centering
  \includegraphics[width=\linewidth]{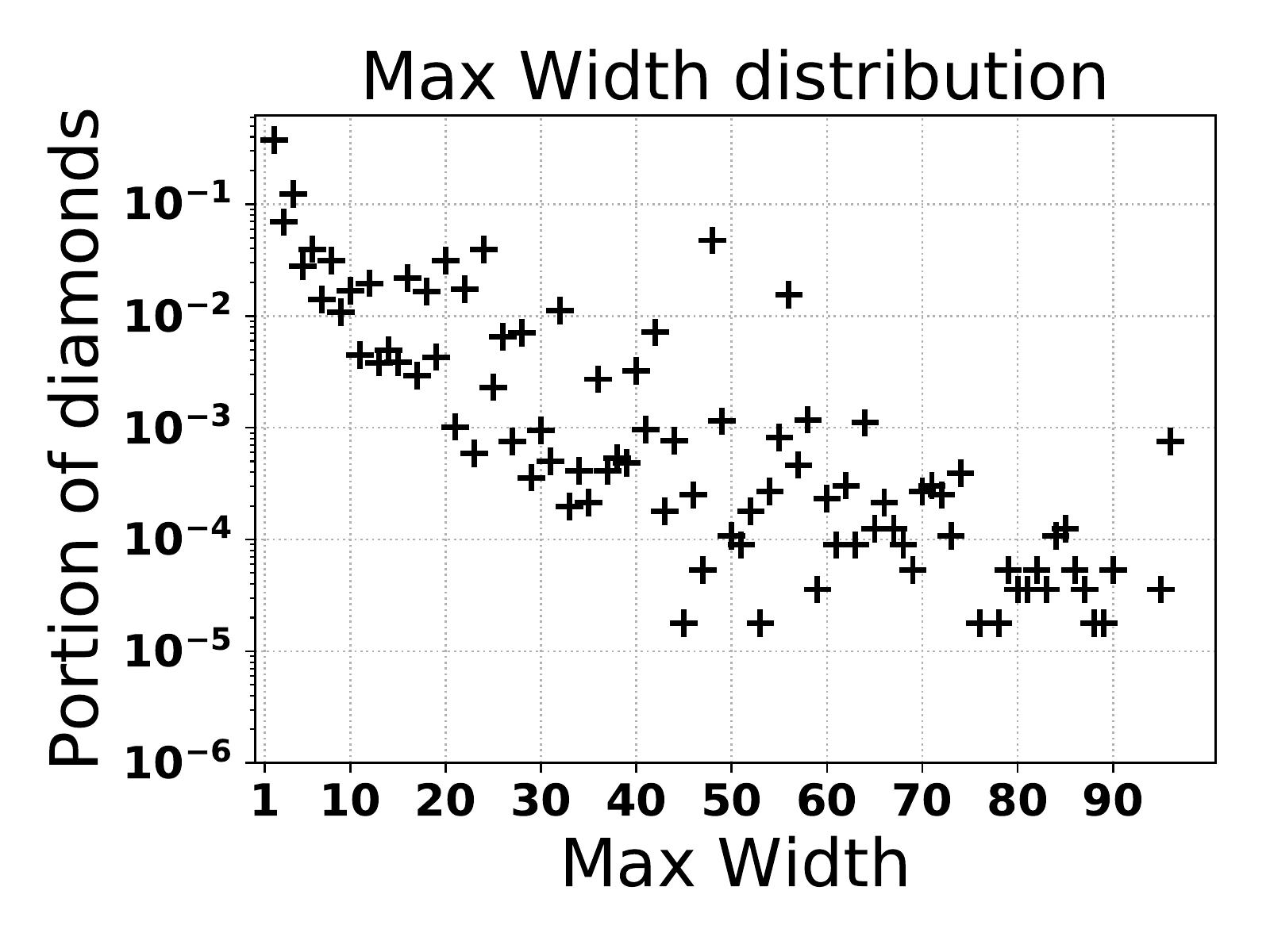}
  \caption{IP level}
\end{subfigure}%
\begin{subfigure}{.25\textwidth}
  \centering
  \includegraphics[width=\linewidth]{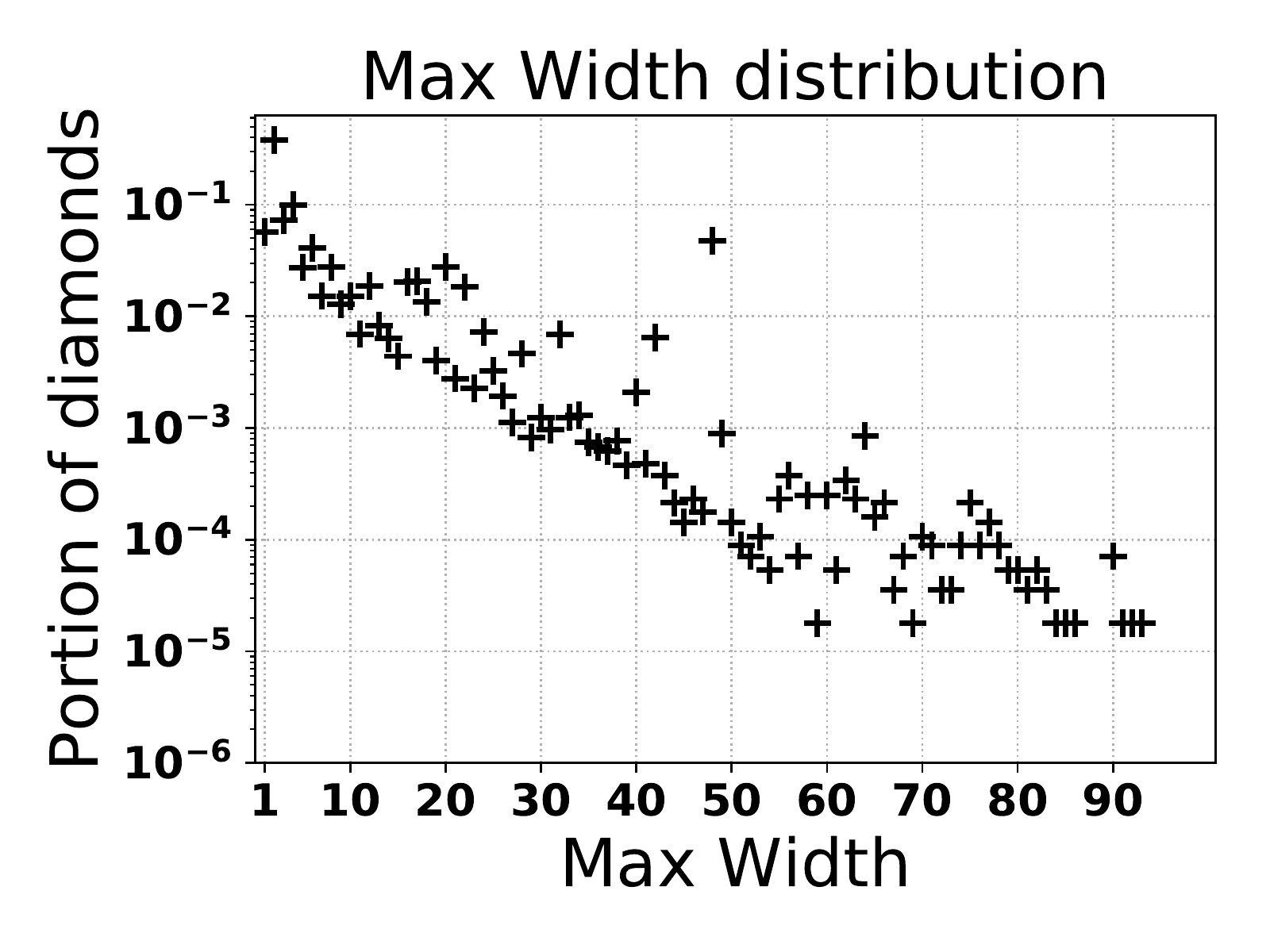}
  \caption{Router level}
\end{subfigure}
\caption{Maximum width of unique diamonds}
\label{fig:metrics-after-alias}
\end{figure}

We looked at the effect of alias resolution on diamond
width. Fig.~\ref{fig:metrics-after-alias} plots the distributions
obtained by the MDA-Lite before and after alias resolution. 
We observe that the peak at maximum width 48 has remained, whereas the
one at 56 has disappeared. On closer inspection, we find that the max
width 56 diamond at the IP level resolved into several smaller
diamonds at the router level. These router-level diamonds were of
unaggregated sizes between 2 and 49 IPs.

\begin{figure}[t]
\begin{center}
    \scalebox{0.7}{  
  \includegraphics[width=0.4\textwidth]{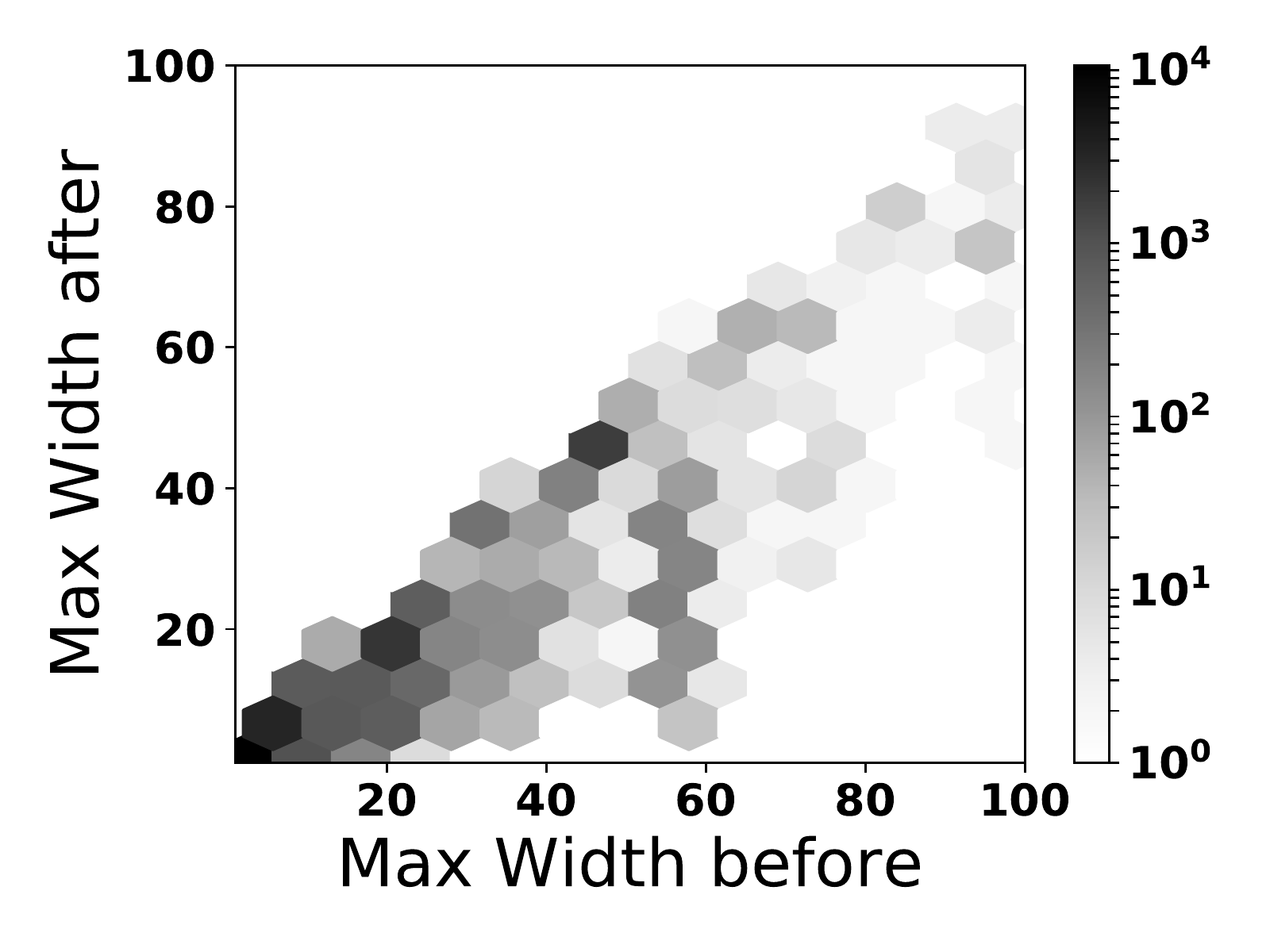}}
\label{fig:compare-before-after-alias-max-width}
\vspace{-1em}
\caption{Joint distribution of maximum width before and after alias resolution}
\label{fig:diamond-to-diamond-comparison}
\end{center}
\vspace{-1em}
\end{figure}

Finally, we looked at width reduction diamond by
diamond. Fig.~\ref{fig:diamond-to-diamond-comparison} plots the joint
distribution of maximum width before and after alias resolution of
those diamonds that changed size. Large width reductions are rare, but
do take place. The darker grey vertical series of values just to the
left of 60 show the maximum width 56 diamonds being broken down into
smaller diamonds at the router level.

\section{Related work}\label{related-work}
\subsection{Contributions}
Our MDA-Lite and multilevel route tracing work builds directly on
Paris
Traceroute~\cite{Augustin:2006:ATA:1177080.1177100,viger2008detection}
and the Multipath Detection Algorithm
(MDA)~\cite{augustin2007measuring,veitch:hal-01298261}. 
It also inscribes itself in the line of measurement
work that has sought to improve our ability to trace the IP level
paths that packets take through the Internet, such as Reverse
Traceroute~\cite{katz2010reverse}, which uses the IP Record Route
option to learn IP addresses on the return path taken by probe
replies; Vanaubel et al.'s Network Fingerprinting
technique~\cite{Vanaubel:2013:NFT:2504730.2504761} for examining the
TTLs of probe replies to determine which type of router might have
sent them and, combined with examination of the MPLS label stack that
is received in an ICMP Time Exceeded message, to trace a path's MPLS
tunnels; or Dublin Traceroute~\cite{barbeiro2016dublin}, which uses
Steven Bellovin's technique~\cite{Bellovin:2002:TCN:637201.637243} for
examining the IP ID field of probe replies for NAT box detection in
order to identify NAT boxes on a multipath route. Similarly to the
latter two, our multilevel route tracing technique goes beyond the interface
level to uncover information about the devices through which packets
pass.

Our multilevel route tracing makes use of existing alias resolution
techniques, notably \textsc{Midar}'s~\cite{midar} state of the art Monotonic
Bounds Test (MBT) for comparing overlapping time series of the IP IDs
of probe replies. The MBT itself builds on the pioneering approaches
of Ally~\cite{Spring:2002:MIT:964725.633039} and
RadarGun~\cite{Bender:2008:FAG:1452520.1452560}. We also use Vanaubel
et al.'s Network
Fingerprinting~\cite{Vanaubel:2013:NFT:2504730.2504761}. But there are
other alias resolution techniques that we do not use. For instance,
the Mercator~\cite{govindan2000heuristics} and
\texttt{iffinder}~\cite{huffaker2002iffinder} approach, which is based on
Pansiot and Grad's technique~\cite{Pansiot:1998:RMT:280549.280555} of
seeing whether a probe to one IP address elicits a reply from
another. Nor do we use Sherry et al.'s prespecified timestamp
technique~\cite{Sherry:2010:RIA:1879141.1879163}. This is because we
currently limit our Traceroute tool to Traceroute-style probing, what
the \textsc{Midar} paper calls ``indirect probing'', that is based
principally on TTL expiry, rather than Ping-style probing, otherwise
called ``direct probing''. But there is no reason in principle,
aside from additional overhead, why such techniques could not be
added. We also do not use Spring et al.'s
technique~\cite{spring2004howto} of examining the names returned by
reverse DNS look-ups and looking for similarities, as this requires
hand-designed rules to reflect each Internet service provider's naming
conventions.
Nor do we use graph analysis based alias
resolution techniques such as
APAR~\cite{Gunes:2009:RIA:1721711.1721715}, \texttt{kapar}~\cite{kapar}, or
DisCarte~\cite{Sherwood:2008:DDI:1402946.1402993}, as these work by
analyzing route traces from multiple sources to multiple
destinations.

Fakeroute is a network simulator purpose-built for one thing:
statistical validation of multipath route detection algorithm implementations
on a variety of topologies. As such,
it does not implement any other features that general network simulators such
as ns-3~\cite{Henderson:2006:NPG:1190455.1190468}, or emulators, such
as GNS3~\cite{gns3}, that can run real router OSes, might offer.

The surveys follow on previous surveys of load balanced paths in the
Internet. Our survey provides an update on Augustin~\etal's
survey~\cite{Augustin:2011:MMR:2042972.2042989} from ten years
ago. Like in Marchetta~\etal's survey~\cite{marchetta2016and}, our
survey transforms IP level traces into router-level topologies, but it
does so with a single tool, while tracing, rather than with additional
measurements by other tools a posteriori. Almeida~\etal characterized
multipath routes in the IPv6
Internet~\cite{almeida2017characterization}, while we have yet to
extend our tool to do the same. Marchetta~\etal's and Almeida~\etal's
surveys are quite recent, from 2016 and 2017 respectively, but they
report only a maximum of 16 interfaces at a given hop, whereas our
survey reveals up to 96.

\subsection{Multipath route tracing on RIPE Atlas}\label{ripe-mda}
RIPE Atlas, because of the resource constraints on its probe boxes,
does not deploy the MDA. Nonetheless, a rudimentary form of MDA can be
realized on the platform.
Within a repeating route tracing measurement, a level of flow ID
variation is permitted: up to 64 variations of what is termed the \textit{Paris ID}.
Measurements are
generally scheduled conservatively, and so over the course of minutes or hours
a probe box may cycle through 64 distinct Paris IDs in its traces towards a destination.
This approach implies that RIPE Atlas is capable of discerning
multiple forward paths between a source and a destination. However, it
does so in a manner that is not optimized, neither in terms of probe
savings nor in terms of statistical guarantees. This has motivated our
search for an improved reduced-overhead MDA.
\section{Conclusion and Future Work}
This paper made four contributions related to Paris Traceroute, each
of which can be developed further. (1) For MDA-Lite, the alternative to the
Multipath Discovery Algorithm (MDA) that significantly reduces overhead
while maintaining a low failure probability, we hope to deepen the
mathematical analysis in order to determine significance levels for
the results, as had been done for the
MDA~\cite{veitch:hal-01298261}. Also, the assumption of uniform load
balancing, which we believe to hold in almost all cases, could be
tested by a rigorous survey, and the algorithm adjusted, which we
believe would be straightforward, to take into account uneven load
balancing if necessary. (2) For Fakeroute, the simulator that enables
validation of a multipath tracing software tool's adherence to its
claimed failure probability bounds, we could extend it to simulate
exceptions to the assumptions made by the MDA and MDA-Lite. Some
assumptions, such as that every probe will receive a reply, often do
not hold in practice. Indeed, ICMP rate limiting is one common cause
of a lack of replies, and a simulator that takes rate limiting into
account could help in designing an algorithm to probe in ways less
likely to trigger rate limiting. Another extension might be to allow
simulation of multilevel route tracing. (3) For multilevel multipath
route tracing, which has provided, a router-level
view of multipath routes, we continue to investigate the differences  
between direct and indirect probing for alias resolution. (4) For the
surveys of multipath routing in the Internet, showing, among other
things, that load balancing topologies have increased in size well
beyond what has been previously reported as recently as 2016, we would
like to repeat them, conducting at a larger scale some of the
side-by-side comparisons of MDA and MDA-Lite that we have so far
conducted only on a smaller scale. Overall, the work is currently
entirely focused on IPv4 and can benefit from being applied to IPv6.

\section*{Acknowledgments}
A research grant from the French Ministry of Defense has made this
work possible. The \textsc{Ant} Lab at USC ISI provided us
with \textsc{Impact} dataset DS-822~\cite{usc-dataset}, which was essential to our study. We
thank: Burim Ljuma, for his precious help in conducting and analyzing
the surveys; earlier team members for their development of MDA Paris
Traceroute and the first version of Fakeroute; our colleagues at
\textsc{Caida}, for their guidance in the use of their tools; and the anonymous
reviewers from both the IMC TPC and the two shadow TPCs, and our shepherd, for their
careful reading of this paper and suggestions for its improvement.
Kevin Vermeulen, Olivier Fourmaux, and Timur Friedman are associated with
Sorbonne Université, CNRS, Laboratoire d'informatique de Paris 6, LIP6,
F-75005 Paris, France. Kevin Vermeulen and Timur Friedman are associated
with the Laboratory of Information, Networking and Communication Sciences, LINCS,
F-75013 Paris, France.
\pagebreak
\bibliographystyle{ACM-Reference-Format}
\bibliography{bibliography}

\end{document}